\documentclass[numbers]{article}

\usepackage[preprint]{arXivPreprint_2025}

\usepackage[utf8]{inputenc} 
\usepackage[T1]{fontenc}    
\usepackage{hyperref}       
\usepackage{url}            
\usepackage{booktabs}       
\usepackage{amsfonts}       
\usepackage{nicefrac}       
\usepackage{microtype}      
\usepackage{xcolor}         

\usepackage{wrapfig}
\usepackage{graphicx}

\usepackage{amsmath}
\usepackage{graphicx}
\usepackage{subcaption}
\usepackage{multirow}
\usepackage{comment}
\usepackage{float}
\usepackage[labelfont=bf]{caption}
\usepackage{natbib}

\title{PSBench: a large-scale benchmark for estimating the accuracy of protein complex structural models}

\author{
  Pawan Neupane\thanks{Equal contribution.} \\
  University of Missouri - Columbia \\
  \texttt{pngkg@missouri.edu} \\
  \And
  Jian Liu\footnotemark[1] \\
  University of Missouri - Columbia\\
  \texttt{jl4mc@missouri.edu} \\
  \And
  Jianlin Cheng\thanks{Corresponding author.} \\
  University of Missouri - Columbia\\
  \texttt{chengji@missouri.edu}
}

\begin{document}

\maketitle

\begin{abstract}

Predicting protein complex structures is essential for protein function analysis, protein design, and drug discovery. While AI methods like AlphaFold can predict accurate structural models for many protein complexes, reliably estimating the quality of these predicted models (estimation of model accuracy, or EMA) for model ranking and selection remains a major challenge. A key barrier to developing effective machine learning-based EMA methods is the lack of large, diverse, and well-annotated datasets for training and evaluation. To address this gap, we introduce PSBench, a benchmark suite comprising four large-scale, labeled datasets generated during the 15th and 16th community-wide Critical Assessment of Protein Structure Prediction (CASP15 and CASP16). PSBench includes over one million structural models covering a wide range of protein sequence lengths, complex stoichiometries, functional classes, and modeling difficulties. Each model is annotated with multiple complementary quality scores at the global, local, and interface levels. PSBench also provides multiple evaluation metrics and baseline EMA methods to facilitate rigorous comparisons. To demonstrate PSBench’s utility, we trained and evaluated GATE, a graph transformer-based EMA method, on the CASP15 data. GATE was blindly tested in CASP16 (2024), where it ranked among the top-performing EMA methods. These results highlight PSBench as a valuable resource for advancing EMA research in protein complex modeling. PSBench is publicly available at: \url{https://github.com/BioinfoMachineLearning/PSBench}.

\end{abstract}

\section{Introduction}

Proteins are essential biological macromolecules whose diverse functions in living organisms are dictated by their three-dimensional (3D) structures. Although experimental techniques such as X-ray crystallography, cryo-electron microscopy (cryo-EM), and nuclear magnetic resonance (NMR) spectroscopy can determine protein structures with high accuracy, these approaches are time-consuming and resource-intensive and can only be applied to a tiny portion ($<$ 0.1\%) of proteins.

To overcome these challenges, machine learning methods\cite{jumper2021highly, baek2021accurate, eickholt2012predicting, guo2021improving, liu2022improving, guo2022prediction, liu2023improving, liu2023enhancing, liu2025improving} for predicting protein structures from sequences have become essential. Among these, AlphaFold\cite{jumper2021highly, evans2021protein, abramson2024accurate} has revolutionized the field by achieving experimental accuracy for predicting the tertiary structures of almost all single-chain proteins (monomers) first and then delivering high-accuracy structure prediction for a large portion of multi-chain protein complexes (multimers). As monomer structure prediction is largely considered solved, protein complex structure prediction is currently one major focus in the field. Despite its success, a critical limitation persists: AlphaFold’s self-estimated model accuracy (quality) scores (e.g., plDDT, pTM, ipTM, and confidence scores) are not always reliable for identifying high-quality predicted complex structures (structural models)\cite{chen20233d}. For instance, using AlphaFold2-Multimer or AlphaFold3 to predict many (e.g., thousands of) structural models for a protein complex target can substantially increase the likelihood of generating some high-quality ones\cite{liu2025improving}, but AlphaFold's own confidence scores or ranking scores often cannot rank them at the top when the ratio of high quality models versus low-quality models is low. As a result, selecting structural models of high quality from a pool of models generated by AlphaFold or other AI methods is a major challenge in protein complex structure prediction and sometimes even harder for users than model generation itself\cite{liu2023enhancing}.

This challenge highlights the importance of Estimation of Model Accuracy (EMA) (also called model quality assessment), which predicts how closely a predicted structural model resembles the native (true) structure before the true structure is known. Reliable EMA tools are not only critical for model selection in the prediction phase, but also vital to prioritize accurate structural models for downstream applications such as protein function annotations and drug discovery. To stimulate the development of EMA methods for protein complex structural models, since 2020, CASP has dedicated one competition category to assess EMA methods\cite{kryshtafovych2023new, kryshtafovych2023critical}. However, EMA methods remain underdeveloped due to two critical gaps: the lack of large, high-quality, labeled complex model datasets to train and test machine learning EMA methods (like ImageNet for image processing) and the lack of user-friendly standardized benchmarks and automated evaluation tools to assess the performance of EMA methods.

\begin{figure}[htbp]
    \centering
    \begin{subfigure}{\textwidth}
        \caption{Pipeline}
        \centering
        \includegraphics[width=\linewidth]{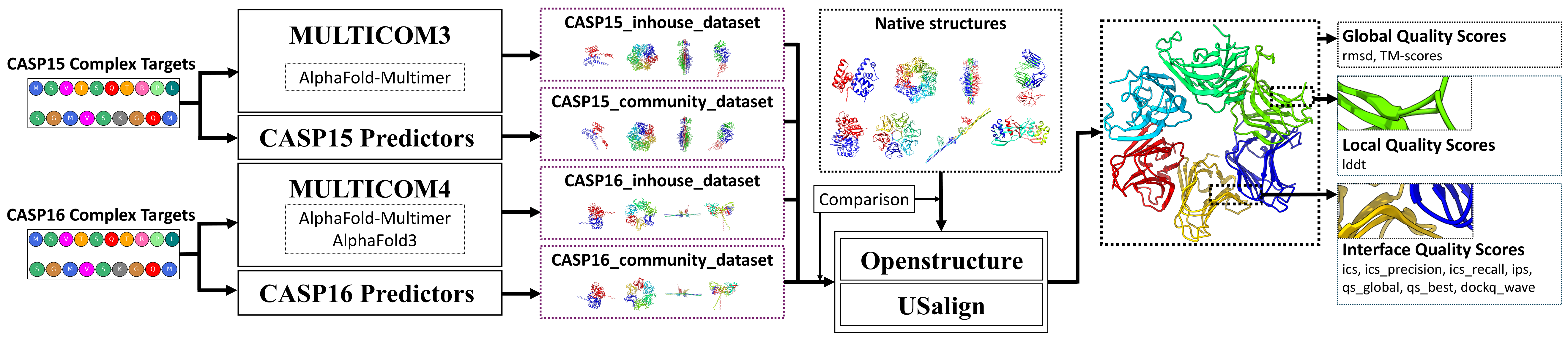}
        \label{fig:pipeline}
    \end{subfigure}
    \hfill
    \begin{subfigure}{0.56\textwidth}
        \caption{Methods}
        \centering
        \includegraphics[width=\linewidth]{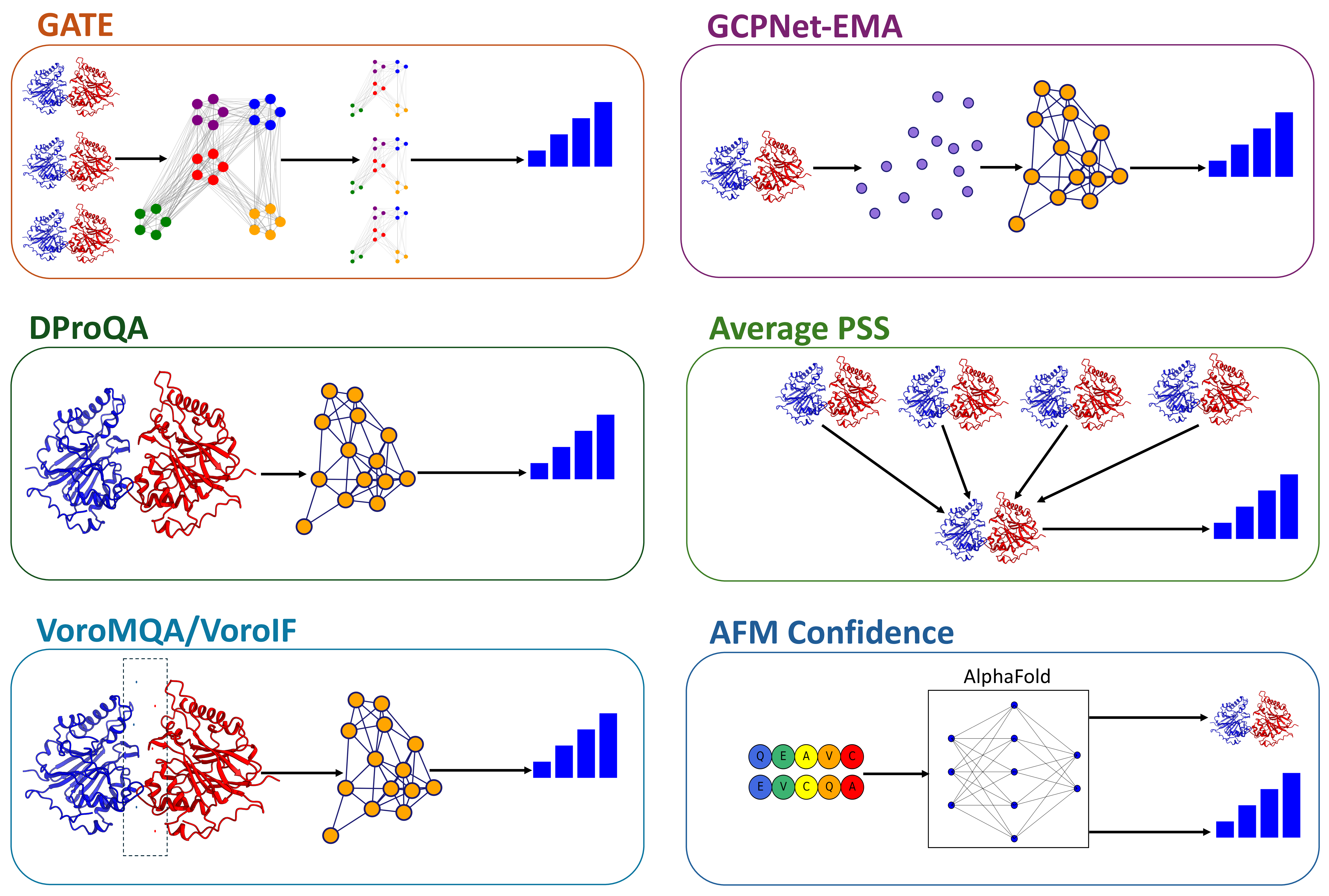}
        \label{fig:methods}
    \end{subfigure}
    \hfill
    \begin{subfigure}{0.42\textwidth}
        \caption{Metrics}
        \centering
        \includegraphics[width=\linewidth]{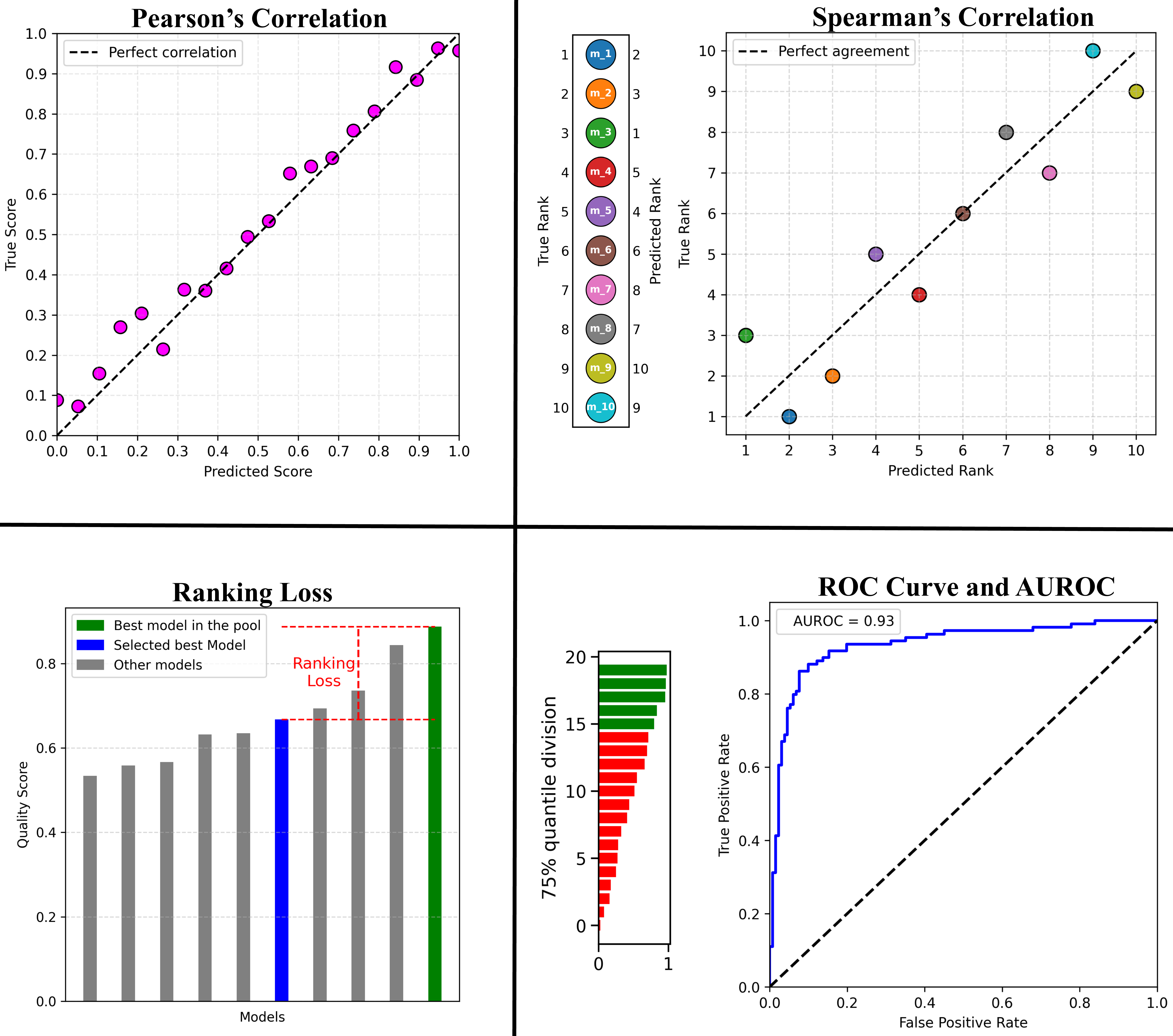}
        \label{fig:metrics}
    \end{subfigure}
    \caption{\textbf{Overview of PSBench.} \textbf{(a) Pipeline.} The PSBench pipeline for preparing four CASP datasets for estimating protein complex model accuracy (EMA). The predicted structural models are compared with native (true) structures to compute global, local, interface quality scores as labels. \textbf{(b) Methods.} Six representative baseline EMA methods for performance comparison. \textbf{(c) Metrics.} Four metrics for evaluating EMA methods: Pearson's correlation, Spearman's correlation, ranking loss, and AUROC (Area Under Receiver Operating Characteristics Curve) for evaluating predicted model quality scores against true ones (labels). The evaluation tools are included in PSBench.}
    \label{fig:PSBench_Introduction}
\end{figure}

To bridge this gap, we introduce PSBench, a comprehensive benchmark for training and testing EMA methods to predict the accuracy (quality) of predicted protein complex structural models and comparing them with baseline methods via multiple complementary metrics (Fig. \ref{fig:PSBench_Introduction}). PSBench consists of four complex structure datasets, including more than one million community-predicted and in-house-predicted structures for protein complex targets of the 2022 CASP15\cite{kryshtafovych2023critical, alexander2023protein, lensink2023impact} and 2024 CASP16 competitions, generated in the truly blind prediction setting (e.g., true structures were unavailable during prediction). The structural models were generated mainly by AlphaFold2-Multimer\cite{evans2021protein} and AlphaFold3\cite{abramson2024accurate} for 79 diverse, representative protein complex targets with different lengths, difficulties and stoichiometries (count of each unique chain in a protein complex), carefully selected by protein structure experts\cite{alexander2023protein}. Each model is rigorously labeled with 10 distinct quality scores spanning global, local and interface accuracy measures (Fig. ~\ref{fig:pipeline}). Importantly, CASP15 models and CASP16 models were generated two years apart and therefore can provide a rigorous split of data for training and testing EMA methods, preventing information leakage and mirroring real-world EMA workflows.

To demonstrate PSBench can be used to train advanced EMA methods and rigorously benchmark them prior to their use,  we trained and tested GATE\cite{liu2025estimating}, a graph transformer-based EMA method on two CASP15 datasets (CASP15\_inhouse\_dataset and CASP15\_community\_dataset) respectively for two purpose: (1) estimating the accuracy of structural models generated by one predictor for model selection and ranking, a typical setting for structure predictors and users; and (2) estimating the accuracy of structural models generated by many predictors in a community, which is a typical setting of CASP EMA competition. We then blindly tested two GATE variants in the blind CASP16 competition held from May to August 2024. 

In the official CASP16 EMA competition category, GATE ranked among the best methods out of 38 participating EMA predictors. In the blind ranking and selection of in-house structural models predicted by our own protein structure system (MULTICOM4) built on top of AlphaFold2-Multimer and AlphaFold3 during CASP16, GATE also outperformed five standard EMA methods and helped MULTICOM4 rank among top predictors in protein complex structure prediction in CASP16\cite{liu2025improving}.
The results demonstrate that PSBench is a valuable resource, including labeled datasets, a model annotation pipeline, baseline EMA methods, and evaluation tools/metrics, for the AI community to develop and benchmark cutting-edge machine learning methods to estimate protein complex model accuracy, addressing a significant bottleneck in the field of protein structure prediction.




\vspace{-2mm}
\section{Background and Related Work}


The development of EMA methods, either physics-/statistical potential-based methods\cite{zhou2002distance, shen2006statistical} or data-driven machine learning methods\cite{cao2017qacon, uziela2016proq2, geng2020iscore}, requires the availability of high-quality datasets of predicted protein complex structural models with annotated quality scores. Particularly, the recently emerged, more powerful deep learning-based EMA methods need to be trained and tested on large, diverse structural model datasets to reliably predict the accuracy (quality) of structural models sampled from the vast protein structure space.

Early benchmark datasets, such as the Docking Benchmark (BM)\cite{vreven2015updates}, PPI4DOCK\cite{yu2016ppi4dock}, and DockGround\cite{kotthoff2022dockground}, consist of structural models generated by traditional protein docking tools like SwarmDock\cite{torchala2013swarmdock}, ZDock\cite{pierce2014zdock}, and pyDock\cite{cheng2007pydock}. However, as docking tools are not accurate and have been largely replaced by much more accurate AlphaFold, these datasets are not suitable for training machine learning methods to predict the quality of structural models generated by widely used AlphaFold. Moreover, these datasets are relatively small. For instance, PPI4DOCK has 54,000 structural models, and  CAPRI Score\_set\cite{lensink2014score_set} comprises 19,013 structural models for 15 targets. Furthermore, the structural models were mostly predicted for small protein complexes (such as homo- and hetero-dimers), which cannot represent large protein complexes consisting of many chains and having more complicated stoichiometries.  

Recent efforts have sought to apply state-of-the-art protein complex structure predictors like AlphaFold2 and AlphaFold2-Multimer to create benchmark datasets. For instance, Multimer-AF2 Dataset (MAF2)\cite{chen2023gated} comprises 9,251 structural models generated by AlphaFold2 and AlphaFold2-Multimer, while Heterodimer-AF2 Dataset (HAF2)\cite{chen2023gated} is a collection of 1,849 structural models for  13 heterodimer proteins generated by the same tools. These datasets have enabled the development of advanced EMA methods such as DProQA\cite{chen2023gated} and ComplexQA\cite{zhang2023complexqa}. However, like the early datasets, these datasets contain a small number of structural models generated for small to medium protein complexes (e.g., sequence length less than 1500 residues), and therefore cannot represent the diverse protein complex structure space well. Moreover, both early and recent benchmark datasets have a limited set of quality scores assigned to each structural model without capturing different aspects of model quality. And the structural models in these datasets were generated in a simulated prediction environment where true structures were known, which is different from the truly blind prediction setting where structural models are predicted prior to true structures being available.

Finally, the previous benchmarks do not provide automated evaluation tools and baseline methods to benchmark new EMA methods, which are important for speeding up the development of machine learning EMA methods.  And they do not include tools to automatically annotate and incorporate new structural models so that they cannot be expanded.

To address the gaps above, we develop PSBench, a large, comprehensive benchmark for developing, training, and testing EMA methods for protein complex. PSBench makes the following unique contributions to the field. 
\begin{itemize}
  \item Providing more than one million complex structural models generated by state-of-the-art deep learning methods (mostly AlphaFold2-Multimer and AlphaFold3), much larger than previous datasets. 
  \item The structural models were generated in the real-world blind prediction setting (CASP15 and CASP16 competitions) without any knowledge of true structures. 
  \item The structural models were generated for 79 diverse protein complex targets carefully selected by protein structure experts, representing 25 different stoichiometries, many function classes, different difficulty levels (easy/medium/hard), and a wide range of sequence lengths (96 to 8460 residues).  
  \item The structural models are assigned 10 complementary quality scores at the local, global, and interface levels, measuring their accuracy from different perspectives, important for training and evaluating EMA methods.
  \item Providing automated evaluation tools for comparing new EMA methods with 6 standard baseline EMA methods, as well as a model annotation (labeling) pipeline for continuous expansion of the datasets. 
  \item{The utility of PSBench for developing state-of-the-art EMA methods was blindly and rigorously proved in CASP16.}
\end{itemize}

\vspace{-2mm}
\section{PSBench Design}
\label{PSBench_design}

PSBench encompasses more than one million predicted structural models distributed in four separate datasets (CASP15\_inhouse\_dataset, CASP15\_community\_dataset, CASP16\_inhouse\_dataset, CASP16\_community\_dataset), which were generated for 79 CASP complex targets during 2022 CASP15 competition and 2024 CASP16 competition (Fig. \ref{fig:PSBench_Introduction}a), respectively. These targets represent 25 distinct stoichiometries (Fig.~\ref{fig:protein_stoichiometries} in Appendix \ref{appendix:diversity_of_targets}) and more than 21 protein classes (Fig.~ \ref{fig:protein_classifications}), providing a broad coverage of protein complexes 


The structural models were compared with the corresponding native (true) structures of the targets by an automated annotation pipeline in PSBench to assign quality scores to them as labels. The annotation pipeline pre-processed structural models and true structures so that they could be aligned and compared by two tools, OpenStructure\cite{biasini2013openstructure, bertoni2017modeling, mariani2013lddt} and USalign\cite{zhang2022us}, generating 10 complementary quality scores measuring model accuracy from three different aspects: global quality, interface quality, and local quality (see the detailed description of these quality scores in Appendix \ref{appendix:quality_score_def}). The global quality scores (variants of \texttt{tm-score} and \texttt{rmsd}) quantify the similarity between the global fold of a model and that of the true structure. The interface quality scores (\texttt{ics},  \texttt{ics\_precision},\texttt{ics\_recall},  \texttt{ips}, \texttt{qs\_global}, \texttt{qs\_best}, and \texttt{dockq\_wave}) measure the quality of interface regions where two chains in a protein complex interact. The local quality score (\texttt{lddt}) measures the accuracy of the location of each residue with respect to its contacted residues. 

The quality scores computed by the annotation pipeline for the models in the CASP15\_community\_dataset and CASP16\_community\_dataset were cross-validated with the scores compiled from the CASP15 and CASP16 websites to make sure that it worked correctly. Users can use one or more quality scores to train and test their EMA methods. Generally, it is recommended at least one global quality score and one interface quality score be used to benchmark EMA methods. The main characteristics of the four datasets are discussed below.

\subsection{CASP15\_inhouse\_dataset}
\begin{figure}[H]
    \centering
    \includegraphics[width=1\linewidth]{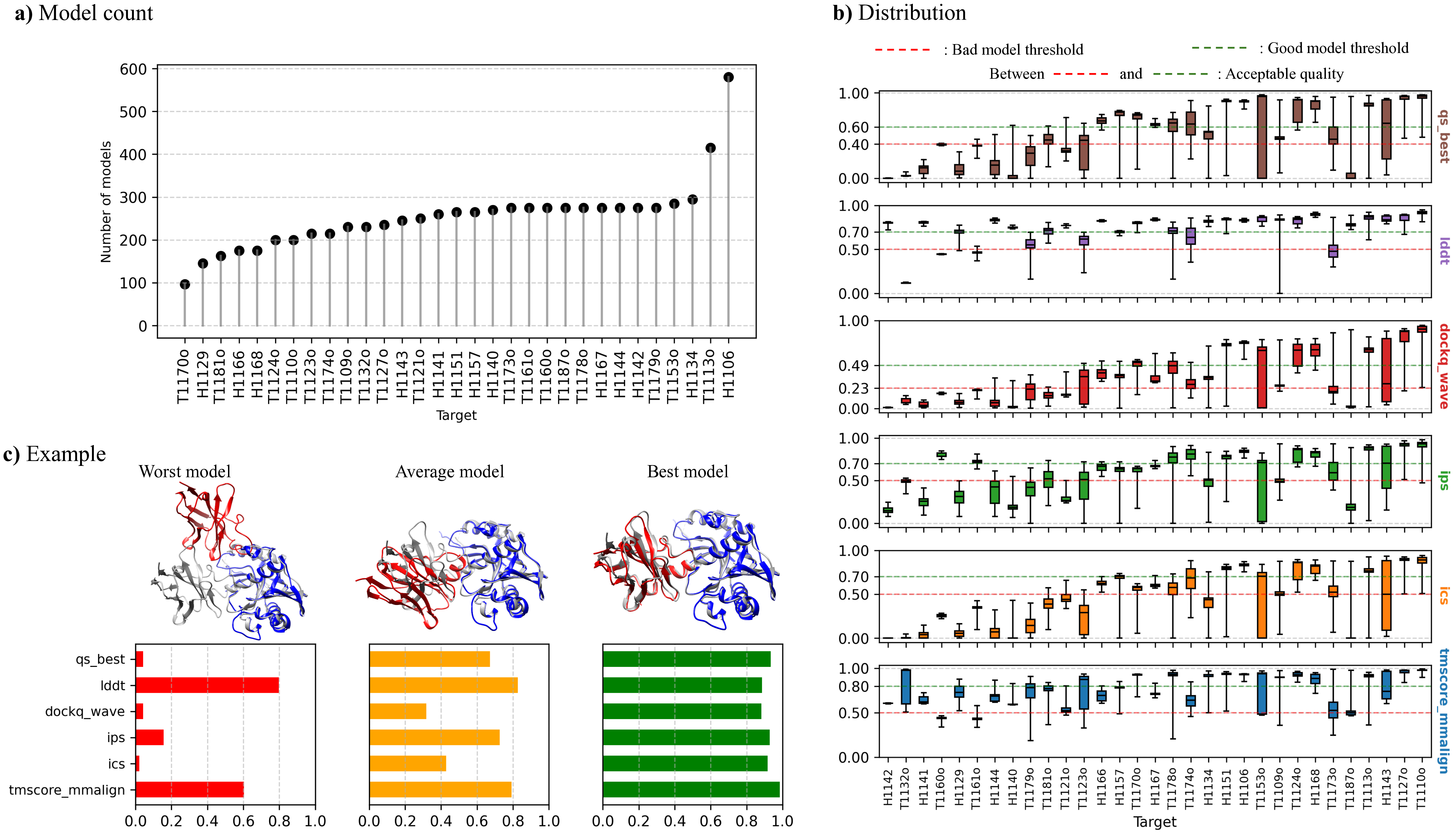}
    \caption{\textbf{CASP15\_inhouse\_dataset.} \textbf{(a) Model count.} Number of models per target in the dataset. \textbf{(b) Score Distribution.} Box plots of each of six representative quality scores of the models for each target. \textbf{(c) Example.} Three representative models (worst, average, best) in terms of sum of the six representative quality scores for a target H1143. Each model with two chains colored in blue and red is superimposed with the true structure in gray.} 
    \label{fig:CASP15_inhouse_dataset}
\end{figure}

The structural models in CASP15\_inhouse\_dataset were generated by our MULTICOM3\cite{liu2023enhancing} system  during the 2022 CASP15 competition. MULTICOM3 generated diverse multiple sequence alignment (MSA), structural templates and hyperparameters as input for AlphaFold2-Multimer\cite{evans2021protein} to predict the structures for 31 CASP15 complex targets and ranked as one of top 10 complex predictors in CASP15. It generated 97 to 580 structural models per target (see model count per target in Fig.~\ref{fig:CASP15_inhouse_dataset}a), resulting in 7,885 models in total included in CASP15\_inhouse\_dataset.  

The 31 CASP15 targets cover a wide range of distinct stoichiometries, sequence lengths, protein classes (see Table \ref{tab:total_models_CASP15_inhouse_dataset} for details). The distribution of six representative quality scores of the structural models for the targets are visualized as box plots in Fig.~\ref{fig:CASP15_inhouse_dataset}b. It can be seen that the targets are of different difficulty levels. Some easy targets have most models above the threshold of good quality, some average targets have one portion of models above the good model threshold and another portion below the bad model threshold, and some difficult targets have most models below the bad model threshold. Therefore, this dataset is ideal for benchmarking EMA methods' capability to work for different kinds of targets and train them to possess the capability. The detailed numbers of good, acceptable, and bad models are reported in Table \ref{tab:dockq_summary_CASP15_inhouse_dataset}. Fig.~\ref{fig:CASP15_inhouse_dataset}c illustrates three representative models (worst, average, and best) of a target H1143 and their quality scores. 

\vspace{-4mm}
Because the structural models in CASP15\_inhouse\_dataset were generated by our own predictor, in addition to a structural file in the Protein Data Bank (PDB) format, every model has four extra features, i.e., four estimated quality scores assigned by AlphaFold2-Multimer during the model generation process, including AlphaFold2-Multimer confidence score (\texttt{afm\_confidence\_score}), interface predicted Template Modeling score (\texttt{iptm}), number of inter-chain predicted aligned errors (<5 Å) (\texttt{num\_inter\_pae}), and predicted multimer DockQ score (\texttt{mpDockQ}) (see Appendix \ref{appendix:additional_features} for details). This is different from CASP15\_community\_dataset generated by many CASP15 predictors that did not submit these features. It is worth noting that these estimated model quality scores are not true quality scores, but can be used as input features to improve the prediction of true quality scores\cite{liu2025estimating}. 
Therefore, CASP15\_inhouse\_dataset is an excellent resource for training EMA methods to rank and select structural models generated by AlphaFold-based structure predictors. 

\subsection{CASP15\_community\_dataset}
The CASP15\_community\_dataset contains structural models submitted by 87 predictors during the CASP15 competition, where each predictor submitted at most 5 models for each target. Most predictors used different variants of AlphaFold2 and AlphaFold2-Multimer to generate models, while some predictors also used other structure prediction methods such as template-based modeling and protein language model-based modeling (e.g., ESMFold\cite{lin2022language}), resulting in a diverse set of structural models. It contains 215 to 319 structural models per target and 10,942 models in total for 40 complex targets (see model count per target in Fig.~\ref{fig:CASP15_community_dataset}a and Table \ref{tab:total_models_CASP15_community_dataset}). 

Unlike CASP15\_inhouse\_dataset generated in an in-house controlled generation process, CASP15\_community\_dataset has more variability in modeling approaches and model quality. As shown in Fig.~\ref{fig:CASP15_community_dataset}b, the quality scores of the models for many targets spread in a wider range, making the dataset particularly valuable for training and benchmark EMA methods for estimating the accuracy of the models generated by a diverse set of predictors with different performance. Fig.~\ref{fig:CASP15_community_dataset}c illustrates three representative models for a target. Table \ref{tab:dockq_summary_CASP15_community_dataset} reports the number of bad, acceptable and good models for each target. Some targets such as T1160o have only one or a few good models, making them challenging targets for EMA methods to pick good ones. 


\subsection{CASP16\_inhouse\_dataset}
The structural models in CASP16\_inhouse\_dataset were  generated by our MULTICOM4 system\cite{liu2025improving} built on top of both AlphaFold2-Multimer and AlphaFold3\cite{abramson2024accurate} during 2024 CASP16 competition. MULTICOM4 ranked no. 1 in the Phase 0 competition of CASP16 in which the stoichiometry of each target was not provided and needed to be predicted and among top 5 in the Phase 1 competition in which the stoichiometry was provided. MULTICOM4 generated 712 to 78,410 structural models for each target (see model count per target in Fig.~\ref{fig:CASP16_inhouse_dataset}a), resulting in 1,009,050 models for 36 complex targets. The stoichiometry, total number of models, number of AlphaFold3 models, sequence length, and protein class for each target are reported in Table \ref{tab:total_models_CASP16_inhouse_dataset}. The CASP16 targets represent a broad range of stoichiometries, protein classes, and sequence lengths. 



Fig.~\ref{fig:CASP16_inhouse_dataset}b illustrates the distribution of the six representative quality scores of the models across targets. In contrast to CASP15\_inhouse\_dataset, CASP16\_inhouse\_dataset exhibits greater variability in model quality, due to a much larger number of models per target and the use of both AlphaFold2-Multimer and AlphaFold3. Table \ref{tab:dockq_summary_CASP16_inhouse_dataset} reports the number of good, acceptable, and bad models for each target. Some targets have many good models, while other have very few, representing different levels of difficulty. As an example, the worst, average, and best models of target T1235o are shown in Fig.~\ref{fig:CASP16_inhouse_dataset}c. 


Same as CASP15\_inhouse\_dataset, each model in CASP16\_inhouse\_dataset also includes four AlphaFold2-Multimer-like self-estimated quality features. 
Additionally,  AlphaFold3-based models have one more feature,  \texttt{af3\_ranking\_score}. 
These scores can be used as input features for EMA methods to predict the quality of the models. Due to these additional features and a very large number of structural models for a diverse set of protein complex targets, CASP16\_inhouse\_dataset is a large, valuable resource to train and benchmark EMA to estimate the accuracy of structural models predicted by both AlphaFold2-Multimer and AlphaFold3. Particularly, to our knowledge, it contains the largest number of labeled protein complex structural models generated by AlphaFold3 to date. 

\subsection{CASP16\_community\_dataset}

CASP16\_community\_dataset comprises 12,904 models for 39 complex targets (166 to 377 models per target) generated by 82 predictors during the CASP16 competition. Most of the predictors used AlphaFold2-Multimer and/or AlphaFold3 to generate structural models, even though some predictors used additional prediction techniques. The per-target model counts, distribution of quality scores and three model examples are illustrated in  Fig.~\ref{fig:CASP16_community_dataset}. The stoichiometry, protein class, sequence length, and number of good/acceptable/bad models of each target are  reported in Tables \ref{tab:total_models_CASP16_community_dataset} and
\ref{tab:dockq_summary_CASP16_community_dataset}.

Like CASP15\_community\_dataset, the models in  CASP16\_community\_dataset originated from a large number of diverse predictors and therefore have a wide range of quality scores (Fig.~\ref{fig:CASP16_community_dataset}b). The dataset is ideal for training and benchmarking EMA methods for predicting the quality of structural models generated by various modern deep learning-based protein structure prediction methods including, but not limited to AlphaFold2-Multimer and AlphaFold3. 

\vspace{-3mm}

\section{Evaluation Framework}
\label{evaluation_framework}
\vspace{-2mm}

\subsection{Evaluating the Utility of PSBench for Training and Testing EMA Methods}

To assess if PSBench could support the development of machine learning EMA methods,  we trained and validated a graph transformer EMA method (GATE\cite{liu2025estimating}) on CASP15\_inhouse\_dataset and CASP15\_community\_dataset separately to obtain two EMA predictors, i.e., (1) one (referred to as GATE-AFM) for predicting the quality scores of structural models generated by AlphaFold and (2) another (referred to as GATE) for predicting the quality scores of structural models generated by many predictors participating in CASP. The two predictors were blindly tested during the CASP16 competition from May to August 2024 as follows.

First, GATE-AFM was used to predict the quality of the in-house structural models generated by MULTICOM4 and select top ones to submit to CASP16 for complex structure prediction competition. To benchmark how GATE performed, five standard EMA methods with source codes available were blindly run in parallel during CASP16, which included DProQA\cite{chen2023gated}, VoroIF-GNN scores\cite{olechnovivc2023voroif},  GCPNet-EMA\cite{morehead2024protein}, PSS\cite{roy2023combining} and AlphaFold2-Multimer confidence scores (AFM Confidence) (Fig. ~\ref{fig:methods}) (see Appendix \ref{appendix:standard_ema_methods} for details). The model quality scores predicted by GATE-AFM and the five methods were compared with the true quality scores available only after CASP16 concluded in December 2024 (see the results in Section \ref{casp16_inhouse_result}). The six methods are included in PSBench for comparison with future EMA methods. 


Second, GATE directly participated in the EMA competition category of CASP16 to evaluate the complex structural models generated by CASP16 complex structure predictors. GATE was assessed along with 37 EMA predictors participated in the CASP16 EMA competition by CASP16 organizers and assessors (see the results in Section \ref{casp16_community_result}). This assessment is highly rigorous and objective because GATE was evaluated with the best EMA predictors in the field by external experts. 




\vspace{-2mm}

\subsection{Protocols of Training and Validating GATE-AFM and GATE on CASP15 Datasets}
\label{gate_training_validation_protocols}
\vspace{-1mm}

GATE-AFM and GATE use the same graph transformer architecture to predict the global quality scores of structural models.  It takes as input a graph of a set of structural models of a target, in which a node denotes a model and an edge connects two similar models, to predict  the quality score (e.g., TM-score) of each model. The common features for each node shared by both GATE-AFM and GATE are the estimated model quality scores assigned by several EMA methods.
The only difference between GATE-AFM and GATE is that the former uses four additional AlphaFold2-Multimer features (confidence scores, ipTM, number of inter-chain predicted aligned errors and mpDockQ) that are only available in CASP15\_inhouse\_dataset but not in CASP15\_community\_dataset. GATE-AFM and GATE use the same set of edge features, including structural similarity scores between two connected nodes.  

GATE was trained and validated on the CASP15\_community\_dataset via 10-fold cross-validation, which comprises 10,935 models of 40 complex targets, plus 187 models from another target (i.e., T1115o) whose native structure is not publicly available. For this target, we obtained its quality scores from the CASP15's website as labels. For each target, a pairwise similarity graph was constructed for all the models of each target first. 2000 subgraphs containing up to 50 nodes were then sampled from the full graph of each target to train and validate GATE. 
GATE-AFM was trained and validated on the CASP15\_inhouse\_dataset in the similar way. To reduce computational costs, only a subset of CASP15\_inhouse\_dataset was used to train GATE-AFM (see Appendix \ref{appendix:TOP5_datasets}).
The details of training and validation are provided in the Appendix \ref{appendix:detail_training_gate}.  

\vspace{-2mm}

\subsection{Evaluation Metrics}
\label{evaluation_metrics}

PSBench provides four complementary metrics (Fig. ~\ref{fig:metrics}) to evaluate the performance of EMA methods. Pearson’s correlation coefficient measures the linear correlation between predicted quality scores from an EMA method and ground-truth quality scores. Spearman’s correlation coefficient evaluates rank-order consistency between predicted scores and ground-truth scores. Ranking loss directly assesses model selection capability by computing the difference between the ground-truth quality score of the truly best model (highest ground-truth score) and the ground truth score of the no.1 model selected by an EMA method, where lower loss values correspond to better selection and 0 means a prefect selection. Area Under the Receiver Operating Characteristic Curve (AUROC) quantifies binary classification performance by labeling models as high-quality (above the 75\textsuperscript{th} percentile of ground-truth scores) or low-quality otherwise. An AUROC of 1.0 indicates perfect classification, while 0.5 corresponds to random guessing. 
Each metrics is usually calculated for the models of each protein target first and then is averaged over all the targets in a dataset as the performance score for an EMA method. 
The scripts that can automatically calculate these scoring metrics for EMA methods are included in PSBench. 



\vspace{-2mm}

\section{Results}
\label{results}
\vspace{-2mm}

\subsection{Blind Prediction Results of Estimating the Accuracy of CASP16 In-house Models}
\label{casp16_inhouse_result}

During CASP16 competition, GATE-AFM pretrained on a subset of CASP15\_inhouse\_dataset, was blindly applied to our in-house models generated by MULTICOM4. Due to the three-day prediction time constraint, we used it to predict the quality of only hundreds of top-ranked models for each target (i.e., the collection of top 5 models generated by each of dozens of predictors based on AlphaFold2-Multimer and AlphaFold3 in MULTICOM4). Four targets (T1249v1o and T1249v2o, T1294v1o and T1294v2o) that have the same sequence but different conformations are excluded. The performance of GATE and the other EMA methods on the top models for the remaining 32 targets (referred to as CASP16\_inhouse\_TOP5\_dataset, a subset of CASP16\_inhouse\_dataset) was compared in Table~\ref{tab:casp16_inhouse_combined_performance}. 



GATE-AFM outperformed the other methods according to almost all evaluation metrics. In terms of a global quality score - TM-score, GATE-AFM achieved the highest Spearman’s correlation (0.283), the lowest ranking loss (0.102), the best AUROC (0.658), and second highest Pearson's correlation (0.372), indicating superior ranking consistency and classification reliability. In terms of an interface quality score - DockQ\_wave, GATE-AFM again outperformed all other methods, attaining the highest Pearson’s correlation (0.431), the highest Spearman’s correlation (0.322), the lowest ranking loss (0.138), and the highest AUROC (0.662). It performed better than AFM Confidence (the default self-estimated quality score of AlphaFold2-Multimer) in terms of all the metrics, demonstrating its significant value of estimating the quality of AlphaFold-generated structural models. In many cases, the improvement of GATE-AFM over the other methods is significant (see the cases marked with * in Table~\ref{tab:casp16_inhouse_combined_performance}). GATE-AFM's strong capability of selecting good models was one important reason that our MULTICOM4 predictors ranked among top predictors in the CASP16 complex structure prediction category\cite{liu2025improving}.  These results show that PSBench can be used to develop state-of-the-art EMA methods. 


\vspace{-2mm}
\begin{table}[H]
    \caption{Performance of EMA methods in estimating the accuracy of the CASP16 in-house models. Metrics include Pearson's correlation (Corr\textsuperscript{P}), Spearman's correlation (Corr\textsuperscript{S}), ranking loss, and AUROC, reported separately for TM-score and DockQ\_wave. Bold font and underline denote the best and second best results respectively.  Significant difference ($p<0.05$) between GATE-AFM and other methods based on the one-sided Wilcoxon signed-rank test is marked with *.}
    \resizebox{\textwidth}{!}{%
    \begin{tabular}{lllllllll}
        \toprule
        \multirow{2}{*}{Method} & \multicolumn{4}{c}{TM-score} & \multicolumn{4}{c}{DockQ\_wave} \\
        \cmidrule(lr){2-5} \cmidrule(lr){6-9}
        & Corr\textsuperscript{P} $\uparrow$ & Corr\textsuperscript{S} $\uparrow$ & Loss $\downarrow$ & AUROC $\uparrow$ 
        & Corr\textsuperscript{P} $\uparrow$ & Corr\textsuperscript{S} $\uparrow$ & Loss $\downarrow$ & AUROC $\uparrow$ \\
        \midrule
        GATE-AFM               & \underline{0.372} & \textbf{0.283} & \textbf{0.102} & \textbf{0.658} & \textbf{0.431} & \textbf{0.322} & \textbf{0.138} & \textbf{0.662} \\
        AFM Confidence         & 0.259*& 0.143*& \underline{0.106} & 0.597*& 0.252*& 0.114*& \underline{0.151} & 0.593*\\
        PSS                   & \textbf{0.394} & \underline{0.261} & 0.114 & \underline{0.647} & \underline{0.369} & \underline{0.284} & 0.154 & 0.645 \\
        GCPNet-EMA            & 0.360 & 0.249 & 0.135 & 0.643 & 0.355 & 0.264 & 0.169 & \underline{0.648} \\
        VoroMQA-dark          & 0.039*& 0.144 & 0.129 & 0.609 & -0.013*& 0.146*& 0.163 & 0.622 \\
        VoroIF-GNN-pCAD-score & 0.073*& 0.105*& 0.167*& 0.589*& 0.074*& 0.137*& 0.204 & 0.615 \\
        VoroIF-GNN-score      & 0.065*& 0.116*& 0.193*& 0.599*& 0.114*& 0.170*& 0.207*& 0.622 \\
        DProQA                & -0.051*& 0.011*& 0.194*& 0.569*& 0.032*& 0.071*& 0.223*& 0.587*\\
        \bottomrule
    \end{tabular}%
    }
    \label{tab:casp16_inhouse_combined_performance}
\end{table}

\vspace{-2mm}

\subsection{Blind Prediction Results in CASP16 EMA Competition}
\label{casp16_community_result}
\vspace{-2mm}

GATE, pretrained on CASP15\_community\_dataset, participated in 2024 CASP16 EMA competition under the predictor name: MULTICOM\_GATE. We downloaded the EMA prediction results of MULTICOM\_GATE and other 37 CASP16 EMA predictors from the CASP16 website. Two very large targets (H1217 and H1227) were excluded because GATE did not generate predictions for their models due to time constraints during CASP16, resulting in using 37 out of 39 targets in CASP16\_community\_dataset for evaluation. 

\begin{wrapfigure}{r}{0.66\textwidth}
\vspace{-4mm}
  \centering
  \includegraphics[width=0.66\textwidth]{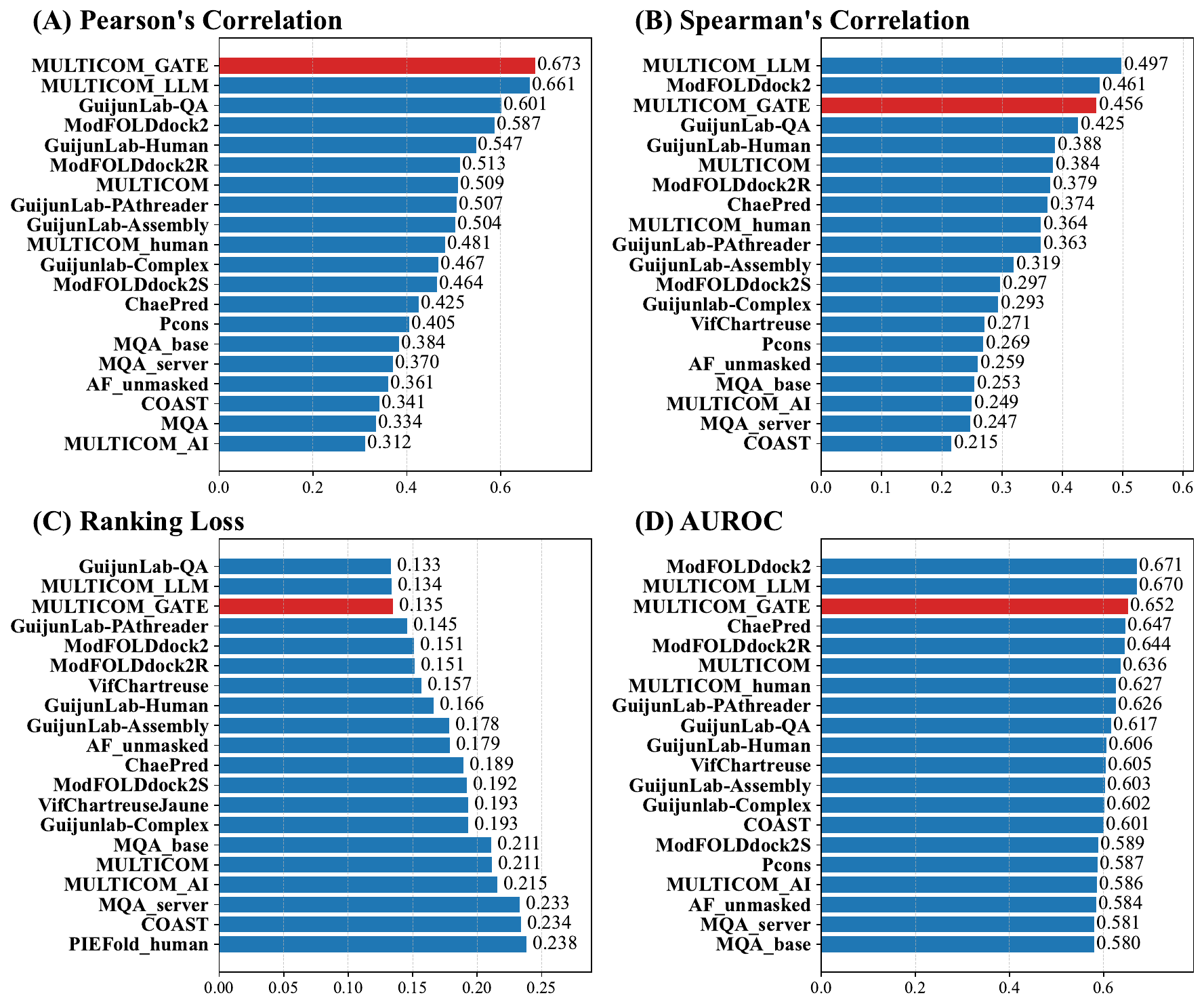}
  \caption{\textbf{CASP16 EMA results.} The performance of top 20 out of 38 CASP16 EMA predictors in predicting TM-scores of structural models of 37 complex targets. \textbf{(a)} Pearson's correlation. \textbf{(b)} Spearman's correlation. \textbf{(c)} Ranking loss. \textbf{(d)} AUROC. MULTICOM\_GATE is highlighted in red. It ranked first, third, third, and third in terms of the four metrics respectively.}
  \label{fig:EMA_TMscore_CASP16}
  \vspace{-4mm}
\end{wrapfigure}

The blind prediction results of top 20 out of 38 CASP16 EMA predictors were shown in Fig.~\ref{fig:EMA_TMscore_CASP16}. In terms of TM-score, MULTICOM\_GATE was ranked first according to Pearson's correlation (0.673), third according to Spearmans' correlation (0.456), ranking loss (0.135) and AUROC (0.652) respectively. Moreover, we assessed the performance of the 38 EMA predictors using a Z-score–based ranking, where MULTICOM\_GATE ranked third (see detailed results in Appendix \ref{appendix:casp16_zscore_ranking}). 
The outstanding performance of MULTICOM\_GATE in CASP16 EMA competition highlights PSBench’s unique value: its high-quality training and test datasets and rigorous evaluation protocols provide a strong framework to develop and validate state-of-the-art machine learning methods.

\vspace{-2mm}
\section{Conclusion, Limitations and Future Work}
\label{conclusion_limitation_and_future_work}
\vspace{-2mm}
PSBench fills critical gaps in protein complex model accuracy estimation by providing a large‐scale benchmark with both data and tools for training and testing EMA methods. It contains more than one million labeled structural models for 79 CASP15/16 targets that cover diverse stoichiometries, function classes, sequence lengths, and difficulty levels. PSBench's value of supporting the development of generalizable EMA methods has been demonstrated by the outstanding performance of GATE trained with it in the blind community-wide CASP16 competition in 2024.  

Despite its strengths, PSBench currently contains structural models for only 79 complexes, which may not fully capture the breadth of protein complexes. To address this limitation, we plan to continuously extend PSBench by incorporating structural models for more targets (e.g., the targets of the upcoming 2026 CASP17 competition and newly released protein complex structures in PDB). We also have provided the model annotation pipeline in PSBench for third-party users to automatically label structural models generated in their research. We will provide an interface to accept structural models contributed by third-parties and acknowledge their contribution in the future release of PSBench. Our goal is to make PSBench a community‐driven resource like ImageNet or MNIST to support AI researchers to solve the critical protein model accuracy estimation problem. 


\vspace{-3mm}
\section{Data and Software Availability}
\label{data_and_software_availability}
\vspace{-1mm}
\textbf{Data Availability}\\
The PSBench datasets are publicly available at Harvard Dataverse: \url{https://dataverse.harvard.edu/dataset.xhtml?persistentId=doi:10.7910/DVN/75SZ1U}. DOI: \url{https://doi.org/10.7910/DVN/75SZ1U} .

\textbf{Software Availability}\\
The programs to evaluate EMA methods on the benchmark datasets and generate labels for new datasets are available at GitHub: \url{https://github.com/BioinfoMachineLearning/PSBench}. The requirements to run PSBench are described in Appendix \ref{appendix:system_requirements}. 


\section{Acknowledgment}
\vspace{-3mm}

We would like to thank CASP organizers and community for sharing CASP15 and CASP16 data and NIH for supporting this research.

\bibliographystyle{plainnat}

\medskip

\medskip

\newpage

\appendix

\renewcommand{\thefigure}{S\arabic{figure}}
\renewcommand{\thetable}{S\arabic{table}}
\setcounter{figure}{0}
\setcounter{table}{0}
\section{Dataset Design}
\label{appendix:dataset_design}

\subsection{Diversity of Protein Complex Targets in PSBench}
\label{appendix:diversity_of_targets}

\begin{figure}[htbp]
    \centering
    \begin{subfigure}{0.46\textwidth}
        \caption{Stoichiometry}
        \centering
        \includegraphics[width=\linewidth]{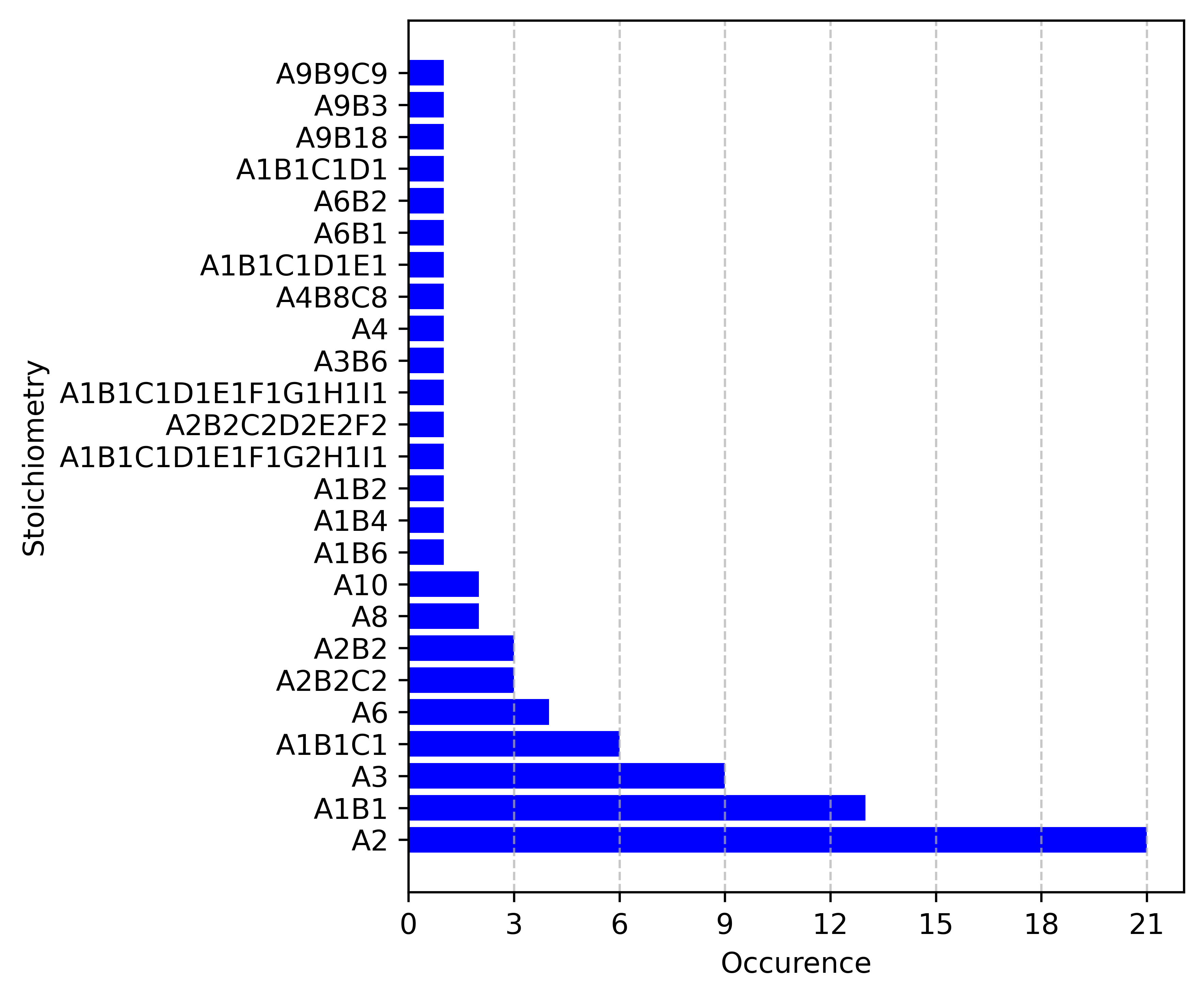}
        \label{fig:protein_stoichiometries}
    \end{subfigure}
    \hfill
    \begin{subfigure}{0.50\textwidth}
        \caption{Protein Class}
        \centering
        \includegraphics[width=\linewidth]{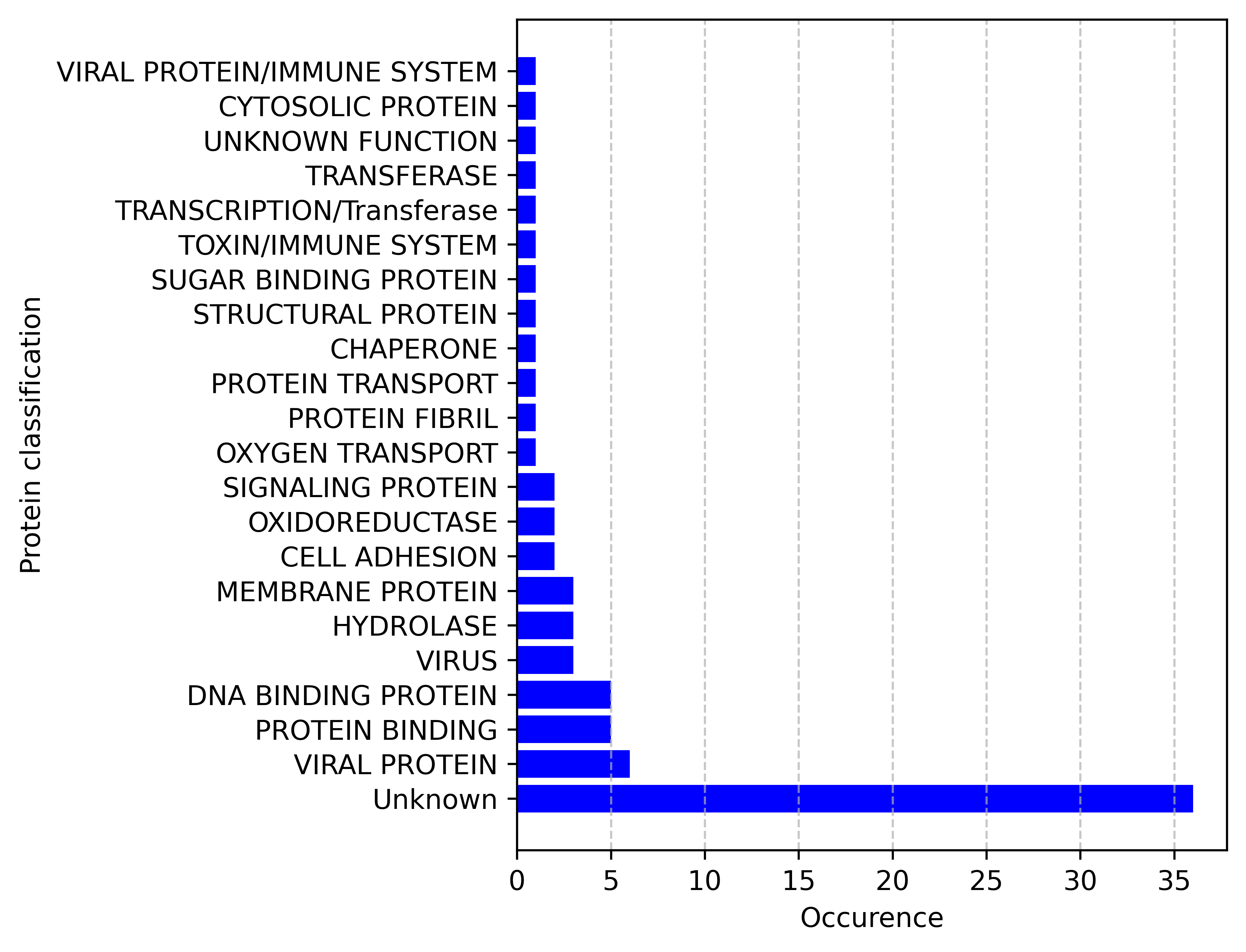}
        \label{fig:protein_classifications}
    \end{subfigure}
    \caption{\textbf{Diversity of 79 protein complex targets.} \textbf{(a)} Number of targets for each of 25 stoichiometries represented in PSBench. A stoichiometry is denoted by letters interleaved with numbers. Each letter represents a unique chain. The number following a letter is the number of the copies (count) of the chain. For instance, A1B2 means a complex has two unique chains A and B, while A has one copy and B has two copies. \textbf{(b)} Number of targets for each of 21 broad protein function classes and an "Unknown" class in PSBench. "Unknown" means there is no class information and therefore may include many different classes.} 
    \label{fig:stoich_proteintype}
\end{figure}

\subsection{Definitions of Quality Scores}
\label{appendix:quality_score_def}
PSBench provides a comprehensive list of quality scores for each protein complex structural model as labels, which are described below. 

\textbf{Global quality scores}
\begin{itemize}
    \item \textbf{tmscore} : The Template Modeling score (TM-score) measures structural similarity between a predicted structure and a reference structure, with higher values (above 0.8) indicating strong agreement. In our evaluation, four variants of the TM-score are used: \textbf{(1) } \texttt{tmscore\_mmalign}, computed using OpenStructure with the USalign plugin and parameters \texttt{-mm 1 -ter 0}, following the CASP16 evaluation protocol; \textbf{(2) }\texttt{tmscore\_usalign}, calculated with the USalign program using parameters \texttt{-ter 1 -TMscore 6}, aligning with the CASP15 evaluation protocol; \textbf{(3) } \texttt{tmscore\_usalign\_aligned}, which further incorporates residue-residue correspondence via an in-house alignment and filtration script before applying USalign with the same parameters; and \textbf{(4) } \texttt{tmscore\_usalign\_aligned\_v0}, an earlier version based on a prior alignment script, used for generating GATE EMA training labels and available only for the CASP15\_inhouse\_dataset. Despite slight procedural differences, all variants apply a consistent threshold interpretation for assessing structural similarity.
    \item \textbf{rmsd} : The Root Mean Square Deviation(\texttt{rmsd}) measures the average distance between corresponding atoms in the model and target structures. It quantifies how much the predicted model deviates from the native structure, with lower values indicating a more accurate structure.

\end{itemize}
\textbf{Local quality score}
\begin{itemize}
    \item \textbf{lddt} : The measures the agreement between inter-atomic distances in the model and the target structure. It evaluates how accurately the overall atomic arrangement is reproduced, focusing on all residues from all chains.
\end{itemize}
\textbf{Interface quality scores}
\begin{itemize}
    \item\textbf{ics} : The Interface Contact Score(\texttt{ics}) measures how accurately predicted residue contacts between protein chains match the true contacts in the target structure. It is a weighted average of F1-scores for each chain-chain interface, where interfaces with more true contacts contribute more to the overall score.
    \item\textbf{ics\_precision} : The Interface Contact Precision(\texttt{ics\_precision}) measures how many predicted residue contacts are correct, focusing on prediction accuracy rather than coverage. It is a weighted average of precision for each chain-chain interface, with larger interfaces (more true contacts) contributing more to the final score.
    \item\textbf{ics\_recall} : The Interface Contact Recall(\texttt{ics\_recall})measures how many true residue contacts are correctly predicted, focusing on coverage. It is a weighted average of recall for each chain-chain interface, with larger interfaces (more true contacts) contributing more to the final score.
    \item\textbf{ips} : The Interface Patch Similarity(\texttt{ips}) measures the similarity between predicted and true interface residue contacts using the Jaccard coefficient. It is a weighted average of the Jaccard index for each chain-chain interface, with larger interfaces (more true contacts) contributing more to the final score.
    \item\textbf{qs\_global} : The QS (global) score(\texttt{qs\_global}) measures the fraction of correctly predicted interface contacts relative to the total number of true or predicted contacts, whichever is larger. It reflects the overall accuracy of contact prediction across the entire protein complex.
    \item\textbf{qs\_best} : The QS (best) score(\texttt{qs\_best}) measures the highest fraction of correctly predicted interface contacts for any single chain-chain interface in the complex. It highlights the best-performing interface prediction within the entire structure.
    \item \textbf{dockq\_wave} : The DockQ\_wave(\texttt{dockq\_wave}) measures the weighted average of DockQ scores across all chain-chain interfaces in the complex. It provides an overall measure of interface prediction quality, combining precision, recall, and Fnat into a single score.
\end{itemize}

\subsection{Additional Input Features for Structural Models in CASP15\_inhouse\_dataset and CASP16\_inhouse\_dataset}
\label{appendix:additional_features}

Each structural model in the four datasets in PSBench is stored as a PDB file, which contains the (x, y, z) coordinates of every atom in the model. In addition, the two in-house datasets (CASP15\_inhouse\_dataset and CASP16\_inhouse\_dataset) and their subsets include the following extra features for each structural model, which can be leveraged by EMA methods.  
\begin{itemize}
    \item \textbf{model\_type} : The type determines whether the model is generated using AlphaFold2 or AlphaFold3. AlphaFold3-based models are only available for CASP16\_inhouse\_dataset and its subset. 
    \item \textbf{afm\_confidence\_score} : The AlphaFold2-Multimer Confidence score (\texttt{afm\_confidence\_score}) determines the confidence score in multimeric protein structures primarily assessed using ipTM and pTM score. For AlphaFold2 program, the AFM confidence score is available upon the completition of the prediction, but for AlphaFold3-based models, since the AFM confidence score is not readily available, it is obtained by the calculation $0.8 \times iptm + 0.2 \times ptm$.
    \item \textbf{af3\_ranking\_score} : The AlphaFold3 Ranking score (\texttt{af3\_ranking\_score}) determines the ranking score as provided by AlphaFold3 program. It is only available for AlphaFold3 generated models in CASP16\_inhouse\_dataset. 
    \item \textbf{iptm} : The Interface Predicted Template Modeling score(\texttt{ipTM}) evaluates the accuracy of the predicted relative positioning of subunits within a protein-protein complex. Scores above 0.8 indicate confident, high-quality predictions, while scores below 0.6 typically reflect failed predictions. Values between 0.6 and 0.8 fall into an intermediate range, where prediction quality is uncertain and may vary.
    \item \textbf{num\_inter\_pae} : Number of inter-chain predicted aligned errors (<5 Å).
    \item \textbf{mpDockQ\cite{bryant2022predicting}/pDockQ\cite{bryant2022improved}} : Multiple-interface predicted DockQ for multimer, or predicted DockQ (pDockQ) for dimer.
\end{itemize}
\clearpage

\subsection{CASP15\_inhouse\_dataset}
\label{appendix:CASP15_inhouse_dataset}

\begin{figure}[H]
    \centering
    \includegraphics[width=1\linewidth]{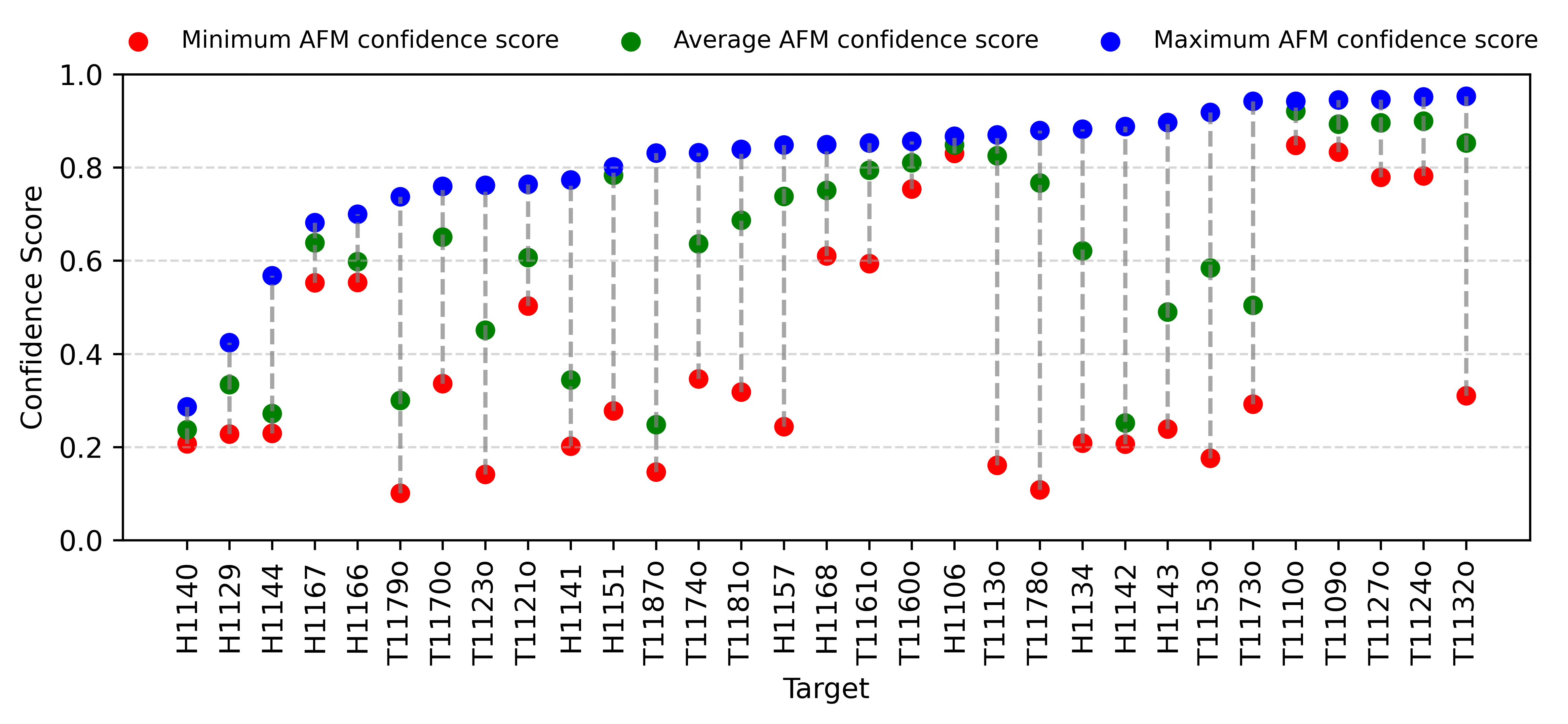}
    \caption{Distribution of AFM confidence scores per target in CASP15\_inhouse\_dataset.}
    \label{fig:CASP15_inhouse_dataset_confidence_score_distribution}
\end{figure}

\begin{table}[H]
\centering
\caption{Summary of target information and number of models per target in CASP15\_inhouse\_dataset. It is worth noting that nine CASP15 targets (H1111, H1114, H1135, H1137, H1171, H1172, H1185, T1176o and T1192o) were excluded because they required alternative structure prediction approaches, such as template-based modeling, due to their large size or the limited number of the predicted structures (e.g., less than 30).}
\label{tab:total_models_CASP15_inhouse_dataset}
\small
\begin{tabular}{lllll}
\toprule
Target & Stoichiometry &    Protein Classification &  Seq. Length &  Total models \\
\midrule
 H1106 &          A1B1 &                 CHAPERONE &              236 &               580 \\
 H1129 &          A1B1 &          MEMBRANE PROTEIN &             1387 &               145 \\
 H1134 &          A1B1 &       TOXIN/IMMUNE SYSTEM &              543 &               295 \\
 H1140 &          A1B1 &           PROTEIN BINDING &              351 &               270 \\
 H1141 &          A1B1 &           PROTEIN BINDING &              346 &               260 \\
 H1142 &          A1B1 &           PROTEIN BINDING &              347 &               275 \\
 H1143 &          A1B1 &           PROTEIN BINDING &              350 &               245 \\
 H1144 &          A1B1 &           PROTEIN BINDING &              341 &               275 \\
 H1151 &          A1B1 & TRANSCRIPTION/Transferase &              228 &               265 \\
 H1157 &          A1B1 &            OXIDOREDUCTASE &             1524 &               265 \\
 H1166 &        A1B1C1 &                   Unknown &              577 &               175 \\
 H1167 &        A1B1C1 &                   Unknown &              560 &               275 \\
 H1168 &        A1B1C1 &                   Unknown &              567 &               175 \\
T1109o &            A2 &                   Unknown &              454 &               230 \\
T1110o &            A2 &                   Unknown &              454 &               200 \\
T1113o &            A2 &             VIRAL PROTEIN &              386 &               415 \\
T1121o &            A2 &       DNA BINDING PROTEIN &              762 &               250 \\
T1123o &            A2 &             VIRAL PROTEIN &              532 &               215 \\
T1124o &            A2 &               TRANSFERASE &              768 &               200 \\
T1127o &            A2 &                   Unknown &              422 &               235 \\
T1132o &            A6 &         CYTOSOLIC PROTEIN &              612 &               230 \\
T1153o &            A2 &                   Unknown &              598 &               285 \\
T1160o &            A2 &       DNA BINDING PROTEIN &               96 &               275 \\
T1161o &            A2 &       DNA BINDING PROTEIN &               96 &               275 \\
T1170o &            A6 &                 HYDROLASE &             1908 &                97 \\
T1173o &            A3 &             CELL ADHESION &              612 &               275 \\
T1174o &            A3 &             CELL ADHESION &             1014 &               215 \\
T1178o &            A2 &             VIRAL PROTEIN &              612 &               275 \\
T1179o &            A2 &             VIRAL PROTEIN &              522 &               275 \\
T1181o &            A3 &                   Unknown &             2064 &               163 \\
T1187o &            A2 &     SUGAR BINDING PROTEIN &              332 &               275 \\
\bottomrule
\end{tabular}
\end{table}

\begin{table}[H]
\centering
\caption{Model quality distribution in terms of dockq\_wave thresholds (bad: score < 0.23, acceptable: 0.23 <= score <0.49, good: 0.49 <= score) for CASP15\_inhouse\_dataset.}
\label{tab:dockq_summary_CASP15_inhouse_dataset}
\small
\begin{tabular}{lccc}
\toprule
target &  Number of bad models &  Number of acceptable models &  Number of good models \\
\midrule
 H1106 &                     0 &                            0 &                    580 \\
 H1129 &                   145 &                            0 &                      0 \\
 H1134 &                    48 &                          231 &                     16 \\
 H1140 &                   269 &                            1 &                      0 \\
 H1141 &                   260 &                            0 &                      0 \\
 H1142 &                   275 &                            0 &                      0 \\
 H1143 &                   112 &                           53 &                     80 \\
 H1144 &                   260 &                           15 &                      0 \\
 H1151 &                     4 &                            0 &                    261 \\
 H1157 &                    58 &                          202 &                      5 \\
 H1166 &                     0 &                          174 &                      1 \\
 H1167 &                     0 &                          215 &                     60 \\
 H1168 &                     0 &                            8 &                    167 \\
T1109o &                     1 &                          189 &                     40 \\
T1110o &                     0 &                            1 &                    199 \\
T1113o &                     9 &                            0 &                    406 \\
T1121o &                   248 &                            2 &                      0 \\
T1123o &                    95 &                          118 &                      2 \\
T1124o &                     0 &                           53 &                    147 \\
T1127o &                     4 &                           12 &                    219 \\
T1132o &                   230 &                            0 &                      0 \\
T1153o &                    98 &                            2 &                    185 \\
T1160o &                   275 &                            0 &                      0 \\
T1161o &                   275 &                            0 &                      0 \\
T1170o &                     1 &                           25 &                     71 \\
T1173o &                   181 &                           67 &                     27 \\
T1174o &                    54 &                          160 &                      1 \\
T1178o &                    25 &                          116 &                    134 \\
T1179o &                   149 &                          126 &                      0 \\
T1181o &                   162 &                            1 &                      0 \\
T1187o &                   270 &                            4 &                      1 \\
\bottomrule
\end{tabular}
\end{table}


\subsection{CASP15\_community\_dataset}
\label{appendix:CASP15_community_dataset}
\begin{figure}[H]
    \centering
    \includegraphics[width=1\linewidth]{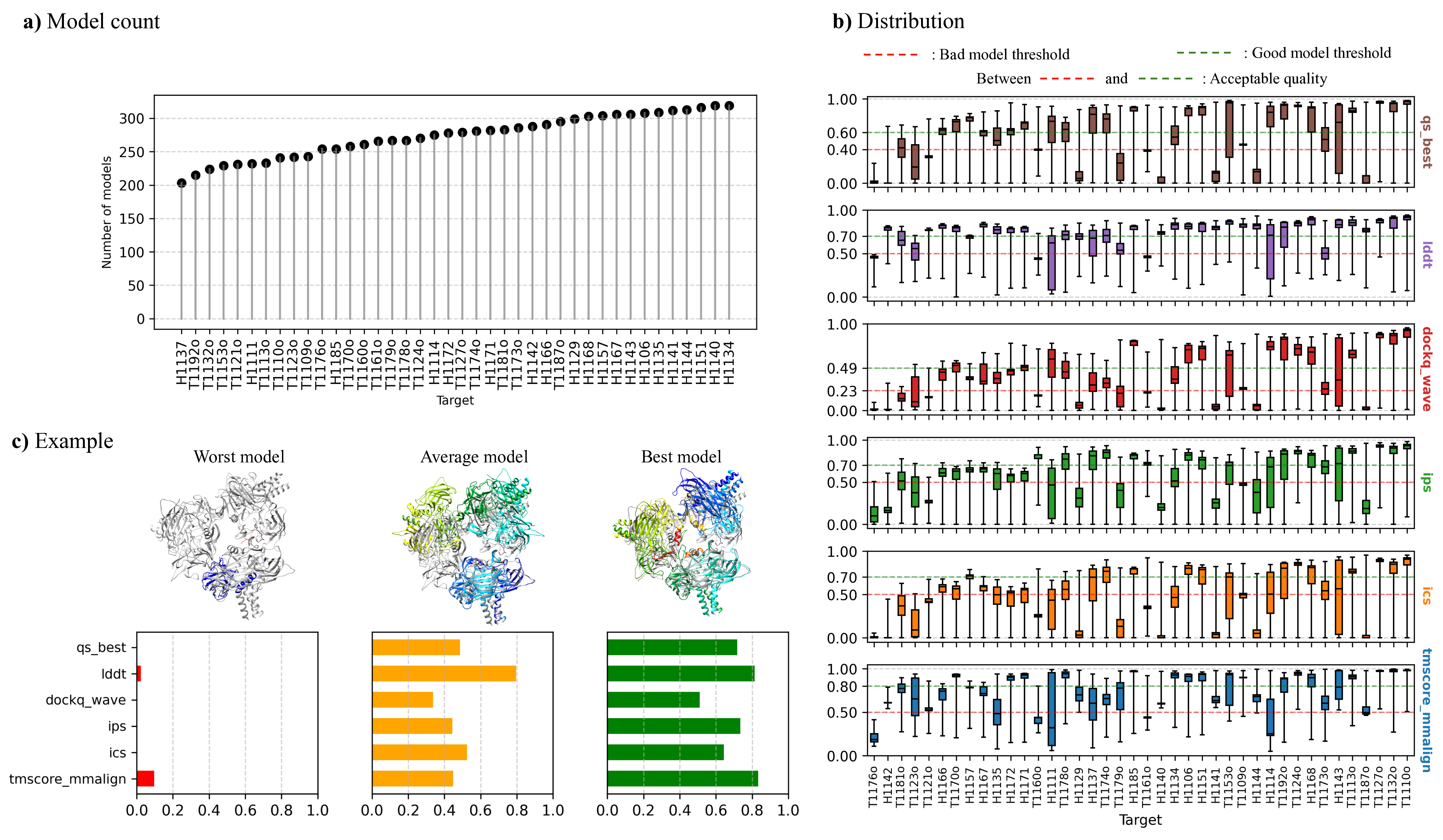}
    
        \caption{\textbf{CASP15\_community\_dataset.} \textbf{(a) Model count.} Number of models per target in the dataset. \textbf{(b) Score Distribution.} Box plots of each of six representative quality scores of the models for each target. \textbf{(c) Example.} Three representative models (worst, average, best) in terms of sum of the six representative quality scores for a target H1135. Each model with individual chains colored is superimposed with the true structure in gray.} 
    \label{fig:CASP15_community_dataset}
\end{figure}
\clearpage

\begin{table}[H]
\centering
\caption{Summary of target information and the number of models per target in  CASP15\_community\_dataset.}
\label{tab:total_models_CASP15_community_dataset}
\small
\begin{tabular}{lllll}
\toprule
Target &      Stoichiometry &    Protein Classification &  Seq. Length &  Total models \\
\midrule
 H1106 &               A1B1 &                 CHAPERONE &              236 &               308 \\
 H1111 &             A9B9C9 &         PROTEIN TRANSPORT &             8460 &               232 \\
 H1114 &             A4B8C8 &            OXIDOREDUCTASE &             7988 &               275 \\
 H1129 &               A1B1 &          MEMBRANE PROTEIN &             1387 &               299 \\
 H1134 &               A1B1 &       TOXIN/IMMUNE SYSTEM &              543 &               319 \\
 H1135 &               A9B3 &        STRUCTURAL PROTEIN &             1830 &               309 \\
 H1137 & A1B1C1D1E1F1G2H1I1 &          MEMBRANE PROTEIN &             4592 &               203 \\
 H1140 &               A1B1 &           PROTEIN BINDING &              351 &               319 \\
 H1141 &               A1B1 &           PROTEIN BINDING &              346 &               312 \\
 H1142 &               A1B1 &           PROTEIN BINDING &              347 &               288 \\
 H1143 &               A1B1 &           PROTEIN BINDING &              350 &               306 \\
 H1144 &               A1B1 &           PROTEIN BINDING &              341 &               313 \\
 H1151 &               A1B1 & TRANSCRIPTION/Transferase &              228 &               316 \\
 H1157 &               A1B1 &            OXIDOREDUCTASE &             1524 &               304 \\
 H1166 &             A1B1C1 &                   Unknown &              577 &               291 \\
 H1167 &             A1B1C1 &                   Unknown &              560 &               306 \\
 H1168 &             A1B1C1 &                   Unknown &              567 &               303 \\
 H1171 &               A6B1 &                 HYDROLASE &             1956 &               282 \\
 H1172 &               A6B2 &                 HYDROLASE &             2004 &               278 \\
 H1185 &           A1B1C1D1 &       DNA BINDING PROTEIN &             1334 &               254 \\
T1109o &                 A2 &                   Unknown &              454 &               243 \\
T1110o &                 A2 &                   Unknown &              454 &               241 \\
T1113o &                 A2 &             VIRAL PROTEIN &              386 &               233 \\
T1121o &                 A2 &       DNA BINDING PROTEIN &              762 &               231 \\
T1123o &                 A2 &             VIRAL PROTEIN &              532 &               242 \\
T1124o &                 A2 &               TRANSFERASE &              768 &               270 \\
T1127o &                 A2 &                   Unknown &              422 &               279 \\
T1132o &                 A6 &         CYTOSOLIC PROTEIN &              612 &               224 \\
T1153o &                 A2 &                   Unknown &              598 &               229 \\
T1160o &                 A2 &       DNA BINDING PROTEIN &               96 &               261 \\
T1161o &                 A2 &       DNA BINDING PROTEIN &               96 &               266 \\
T1170o &                 A6 &                 HYDROLASE &             1908 &               258 \\
T1173o &                 A3 &             CELL ADHESION &              612 &               286 \\
T1174o &                 A3 &             CELL ADHESION &             1014 &               281 \\
T1176o &                 A8 &          UNKNOWN FUNCTION &             1360 &               254 \\
T1178o &                 A2 &             VIRAL PROTEIN &              612 &               267 \\
T1179o &                 A2 &             VIRAL PROTEIN &              522 &               267 \\
T1181o &                 A3 &                   Unknown &             2064 &               283 \\
T1187o &                 A2 &     SUGAR BINDING PROTEIN &              332 &               295 \\
T1192o &                A10 &       DNA BINDING PROTEIN &             4180 &               215 \\
\bottomrule
\end{tabular}
\end{table}

\begin{table}[H]
\centering
\caption{Model quality distribution in terms of dockq\_wave thresholds (bad: score < 0.23, acceptable: 0.23 <= score < 0.49, good: 0.49 <= score) for CASP15\_community\_dataset.}
\label{tab:dockq_summary_CASP15_community_dataset}
\small
\begin{tabular}{lccc}
\toprule
Target &  Number of bad models &  Number of acceptable models &  Number of good models \\
\midrule
 H1106 &                    40 &                            6 &                    262 \\
 H1111 &                    11 &                           72 &                    146 \\
 H1114 &                    20 &                           15 &                    238 \\
 H1129 &                   286 &                            3 &                     10 \\
 H1134 &                    70 &                          164 &                     85 \\
 H1135 &                    40 &                          247 &                     22 \\
 H1137 &                    59 &                           99 &                     45 \\
 H1140 &                   311 &                            6 &                      2 \\
 H1141 &                   300 &                            1 &                     11 \\
 H1142 &                   287 &                            1 &                      0 \\
 H1143 &                   129 &                           40 &                    137 \\
 H1144 &                   292 &                           12 &                      9 \\
 H1151 &                    54 &                           12 &                    250 \\
 H1157 &                    48 &                          222 &                     34 \\
 H1166 &                    16 &                          224 &                     51 \\
 H1167 &                    19 &                          201 &                     86 \\
 H1168 &                    18 &                           39 &                    246 \\
 H1171 &                    24 &                           89 &                    169 \\
 H1172 &                    27 &                          207 &                     44 \\
 H1185 &                    16 &                           11 &                    227 \\
T1109o &                    37 &                          187 &                     19 \\
T1110o &                    10 &                           14 &                    217 \\
T1113o &                    25 &                            7 &                    201 \\
T1121o &                   217 &                           14 &                      0 \\
T1123o &                   155 &                           78 &                      9 \\
T1124o &                    15 &                           18 &                    237 \\
T1127o &                     8 &                            8 &                    263 \\
T1132o &                     9 &                            7 &                    208 \\
T1153o &                    68 &                           16 &                    145 \\
T1160o &                   260 &                            0 &                      1 \\
T1161o &                   259 &                            4 &                      3 \\
T1170o &                    27 &                           61 &                    170 \\
T1173o &                   133 &                          114 &                     39 \\
T1174o &                    44 &                          234 &                      3 \\
T1176o &                   254 &                            0 &                      0 \\
T1178o &                    41 &                          106 &                    120 \\
T1179o &                   171 &                           86 &                     10 \\
T1181o &                   258 &                           25 &                      0 \\
T1187o &                   272 &                            5 &                     18 \\
T1192o &                    25 &                           16 &                    174 \\
\bottomrule
\end{tabular}
\end{table}


\subsection{CASP16\_inhouse\_dataset}
\label{appendix:CASP16_inhouse_dataset}
\begin{figure}[H]
    \centering
    \includegraphics[width=1\linewidth]{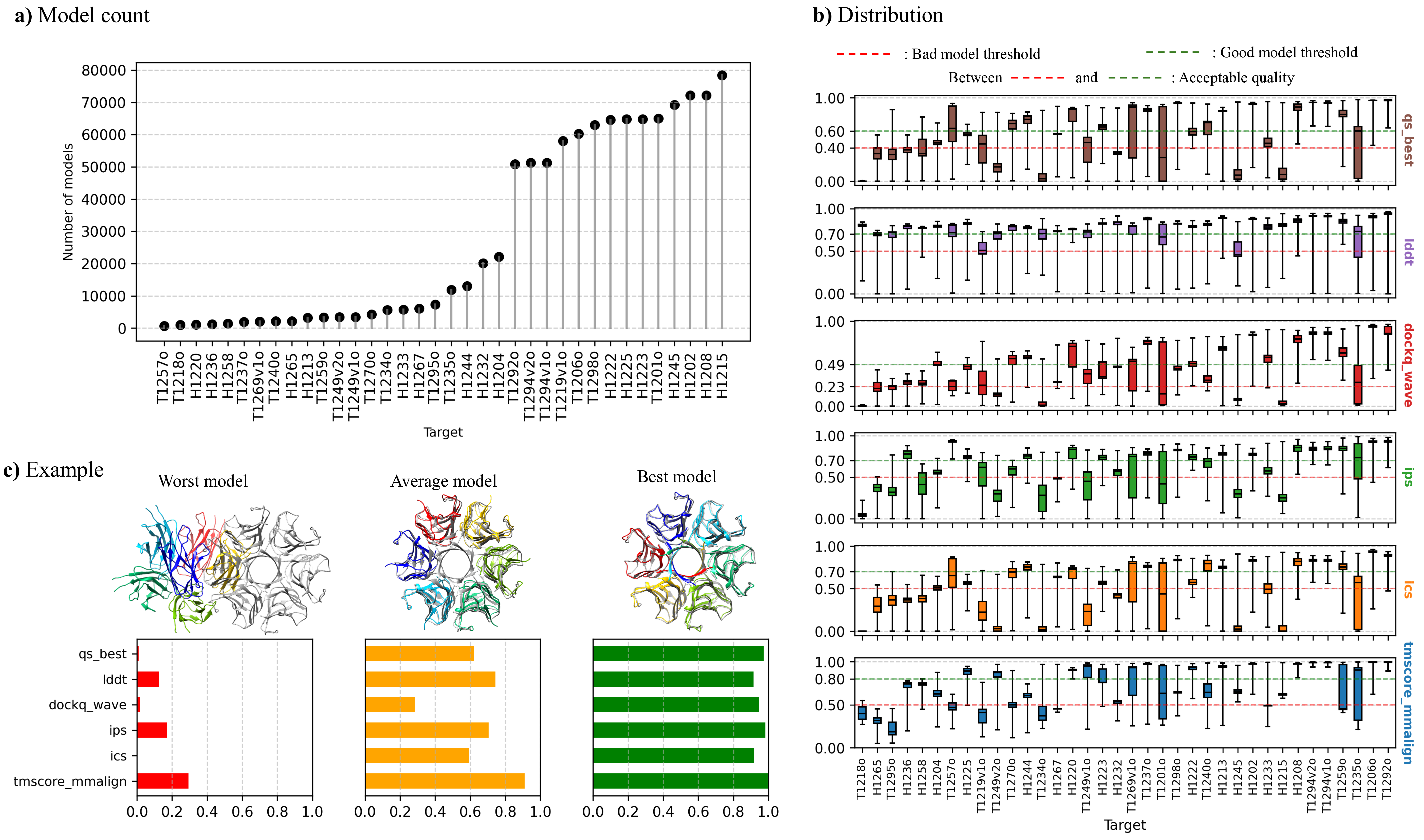}

 \caption{\textbf{CASP16\_inhouse\_dataset.} \textbf{(a) Model count.} Number of models per target in the dataset. \textbf{(b) Score Distribution.} Box plots of each of six representative quality scores of the models for each target. \textbf{(c) Example.} Three representative models (worst, average, best) in terms of sum of the six representative quality scores for a target T1235o. Each model with individual chains colored is superimposed with the true structure in gray.}
    
    \label{fig:CASP16_inhouse_dataset}
\end{figure}

\begin{figure}[H]
    \centering
    \includegraphics[width=1\linewidth]{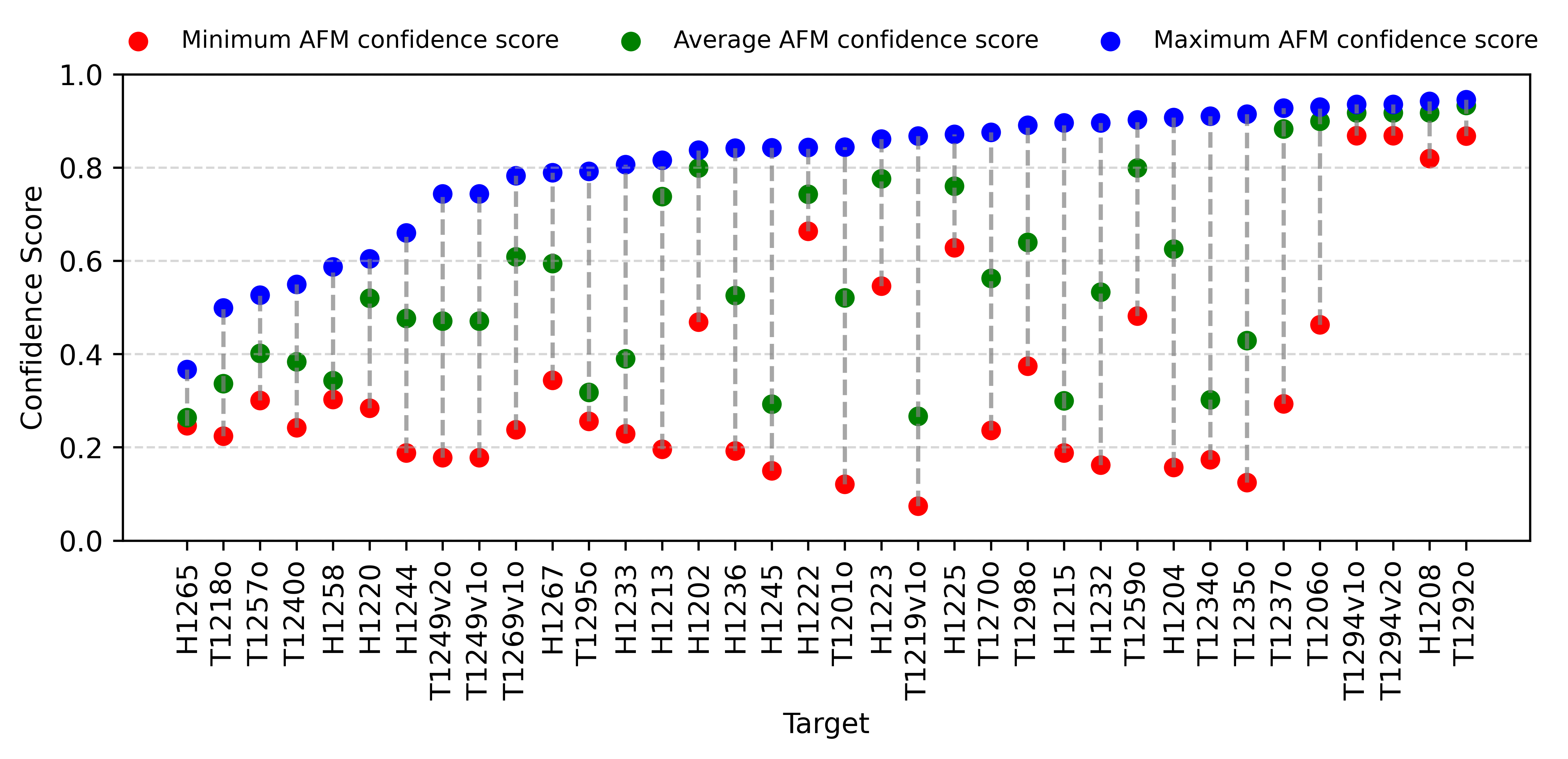}
    \caption{Distribution of AFM confidence scores of the structural models per target in CASP16\_inhouse\_dataset.}
    \label{fig:CASP16_inhouse_dataset_confidence_score_distribution}
\end{figure}

\begin{figure}[H]
    \centering
    \includegraphics[width=1\linewidth]{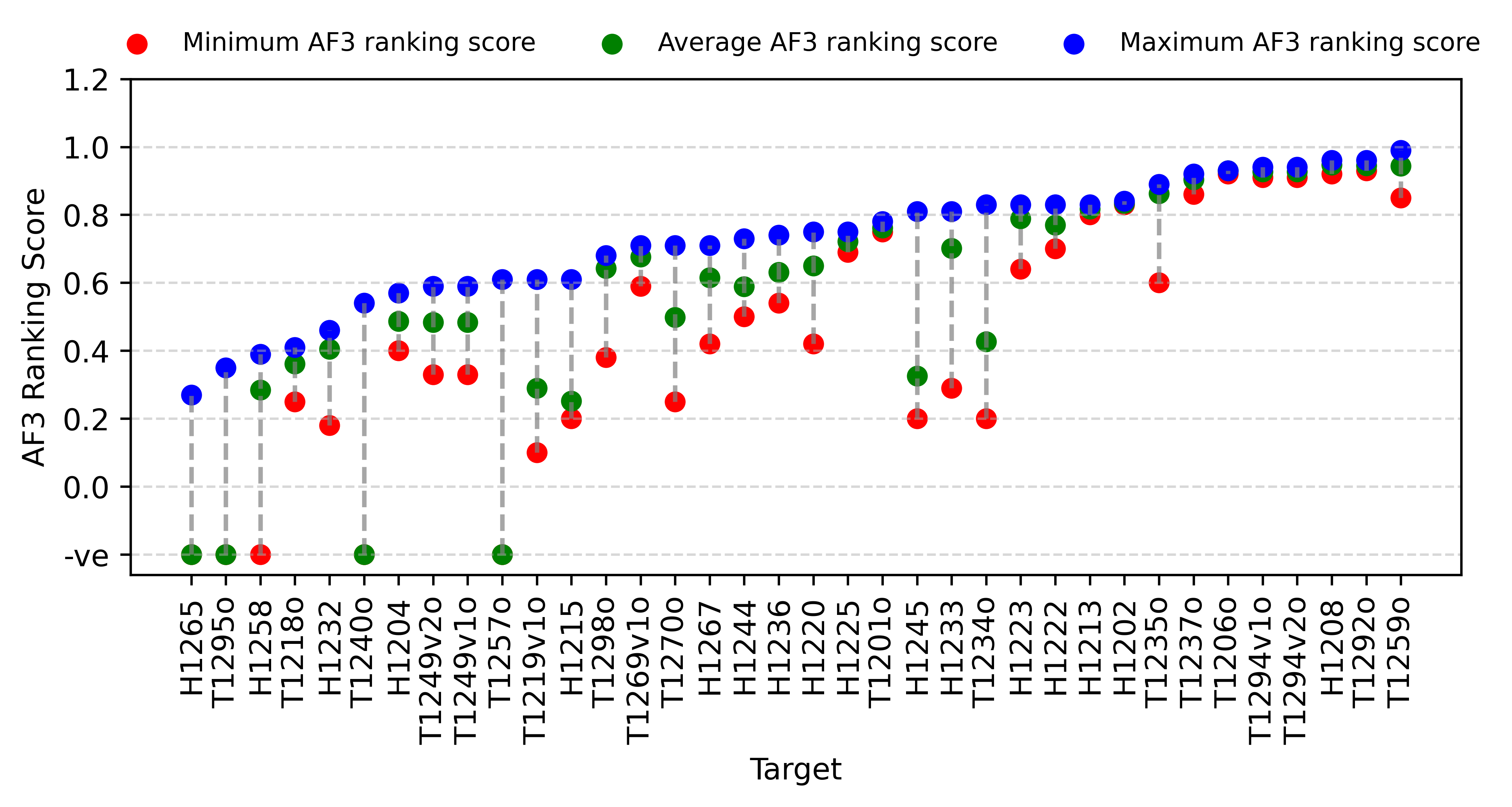}
    \caption{Distribution of AF3 ranking scores of the structural models per target in CASP16\_inhouse\_dataset.}
    \label{fig:CASP16_inhouse_dataset_af3_ranking_score_distribution}
\end{figure}


\begin{table}[H]
\centering
\caption{Summary of target information and number of models per target in CASP16\_inhouse\_dataset. It is worth noting that the length of three CASP16 targets (H1217, H1227, and H1272) exceeded the limit (about 5,000 residues) of running AlphaFold and required other structure prediction techniques such as template-based modeling. To make this dataset include only structural models generated by AlphaFold, they are excluded.}
\label{tab:total_models_CASP16_inhouse_dataset}
\small
\begin{tabular}{lllll}
\toprule
  Target & Stoichiometry &      Protein Classification &  Seq. Length &  Total models \\
\midrule
   H1202 &          A2B2 &           SIGNALING PROTEIN &              380 &             72185 \\
   H1204 &        A2B2C2 &            OXYGEN TRANSPORT &              858 &             22110 \\
   H1208 &          A1B1 &                     Unknown &              646 &             72200 \\
   H1213 &    A1B1C1D1E1 &                     Unknown &             1373 &              3200 \\
   H1215 &          A1B1 &                     Unknown &              369 &             78410 \\
   H1220 &          A1B4 &                     Unknown &             2515 &              1150 \\
   H1222 &        A1B1C1 &                     Unknown &              485 &             64600 \\
   H1223 &        A1B1C1 &                     Unknown &              486 &             64800 \\
   H1225 &        A1B1C1 &                     Unknown &              483 &             64799 \\
   H1232 &          A2B2 &               VIRAL PROTEIN &              924 &             20090 \\
   H1233 &        A2B2C2 & VIRAL PROTEIN/IMMUNE SYSTEM &             1316 &              5700 \\
   H1236 &          A3B6 &                       VIRUS &             1929 &              1178 \\
   H1244 &        A2B2C2 &                     Unknown &              850 &             13000 \\
   H1245 &          A1B1 &                     Unknown &              317 &             69200 \\
   H1258 &          A1B2 &                     Unknown &             3092 &              1423 \\
   H1265 &         A9B18 &                     Unknown &             3924 &              2152 \\
   H1267 &          A2B2 &                     Unknown &             1852 &              6050 \\
  T1201o &            A2 &           SIGNALING PROTEIN &              420 &             65020 \\
  T1206o &            A2 &               VIRAL PROTEIN &              474 &             60205 \\
  T1218o &            A2 &                     Unknown &             2328 &               949 \\
T1219v1o &           A10 &                     Unknown &              320 &             58000 \\
  T1234o &            A3 &                       VIRUS &             1239 &              5600 \\
  T1235o &            A6 &                       VIRUS &              690 &             11900 \\
  T1237o &            A4 &                     Unknown &             1952 &              1970 \\
  T1240o &            A3 &                     Unknown &             1959 &              2125 \\
T1249v1o &            A3 &                     Unknown &             1464 &              3450 \\
T1249v2o &            A3 &                     Unknown &             1464 &              3450 \\
  T1257o &            A3 &                     Unknown &             3789 &               712 \\
  T1259o &            A3 &                     Unknown &              729 &              3350 \\
T1269v1o &            A2 &              PROTEIN FIBRIL &             2820 &              2025 \\
  T1270o &            A6 &                     Unknown &             2622 &              4278 \\
  T1292o &            A2 &                     Unknown &              392 &             50800 \\
T1294v1o &            A2 &                     Unknown &              428 &             51300 \\
T1294v2o &            A2 &                     Unknown &              428 &             51300 \\
  T1295o &            A8 &                     Unknown &             3752 &              7369 \\
  T1298o &            A2 &                     Unknown &              684 &             63000 \\
\bottomrule
\end{tabular}
\end{table}

\begin{table}[H]
\centering
\caption{Model quality distribution in terms of dockq\_wave scores (bad: score < 0.23, acceptable: 0.23 <= score < 0.49, good: 0.49 <= score) for CASP16\_inhouse\_dataset.}
\label{tab:dockq_summary_CASP16_inhouse_dataset}
\small
\begin{tabular}{lccc}
\toprule
  Target &  Number of bad models &  Number of acceptable models &  Number of good models \\
\midrule
   H1202 &                     0 &                          448 &                  71737 \\
   H1204 &                   144 &                        12112 &                   9854 \\
   H1208 &                     0 &                         1135 &                  71065 \\
   H1213 &                   100 &                            9 &                   3091 \\
   H1215 &                 77215 &                         1024 &                    171 \\
   H1220 &                    16 &                          505 &                    629 \\
   H1222 &                     0 &                        25083 &                  39517 \\
   H1223 &                    84 &                        47480 &                  17236 \\
   H1225 &                    20 &                        46979 &                  17800 \\
   H1232 &                    26 &                        18138 &                   1926 \\
   H1233 &                     2 &                          883 &                   4815 \\
   H1236 &                   258 &                          920 &                      0 \\
   H1244 &                     1 &                          400 &                  12599 \\
   H1245 &                 66839 &                         1682 &                    679 \\
   H1258 &                   252 &                         1171 &                      0 \\
   H1265 &                  1279 &                          873 &                      0 \\
   H1267 &                     1 &                         6038 &                     11 \\
  T1201o &                 34545 &                         1227 &                  29248 \\
  T1206o &                     0 &                         1441 &                  58764 \\
  T1218o &                   949 &                            0 &                      0 \\
T1219v1o &                 25852 &                        19207 &                  12941 \\
  T1234o &                  5063 &                          527 &                     10 \\
  T1235o &                  5260 &                         4097 &                   2543 \\
  T1237o &                     2 &                           29 &                   1939 \\
  T1240o &                    80 &                         2009 &                     36 \\
T1249v1o &                   803 &                         2261 &                    386 \\
T1249v2o &                  3255 &                          183 &                     12 \\
  T1257o &                   352 &                          360 &                      0 \\
  T1259o &                     0 &                           19 &                   3331 \\
T1269v1o &                   626 &                          178 &                   1221 \\
  T1270o &                   108 &                          965 &                   3205 \\
  T1292o &                     0 &                          509 &                  50291 \\
T1294v1o &                     0 &                            0 &                  51300 \\
T1294v2o &                     0 &                            0 &                  51300 \\
  T1295o &                  3668 &                         3701 &                      0 \\
  T1298o &                  1204 &                        48835 &                  12961 \\
\bottomrule
\end{tabular}
\end{table}

\subsection{CASP16\_community\_dataset}
\label{appendix:CASP16_community_dataset}
\begin{figure}[H]
    \centering
    \includegraphics[width=1\linewidth]{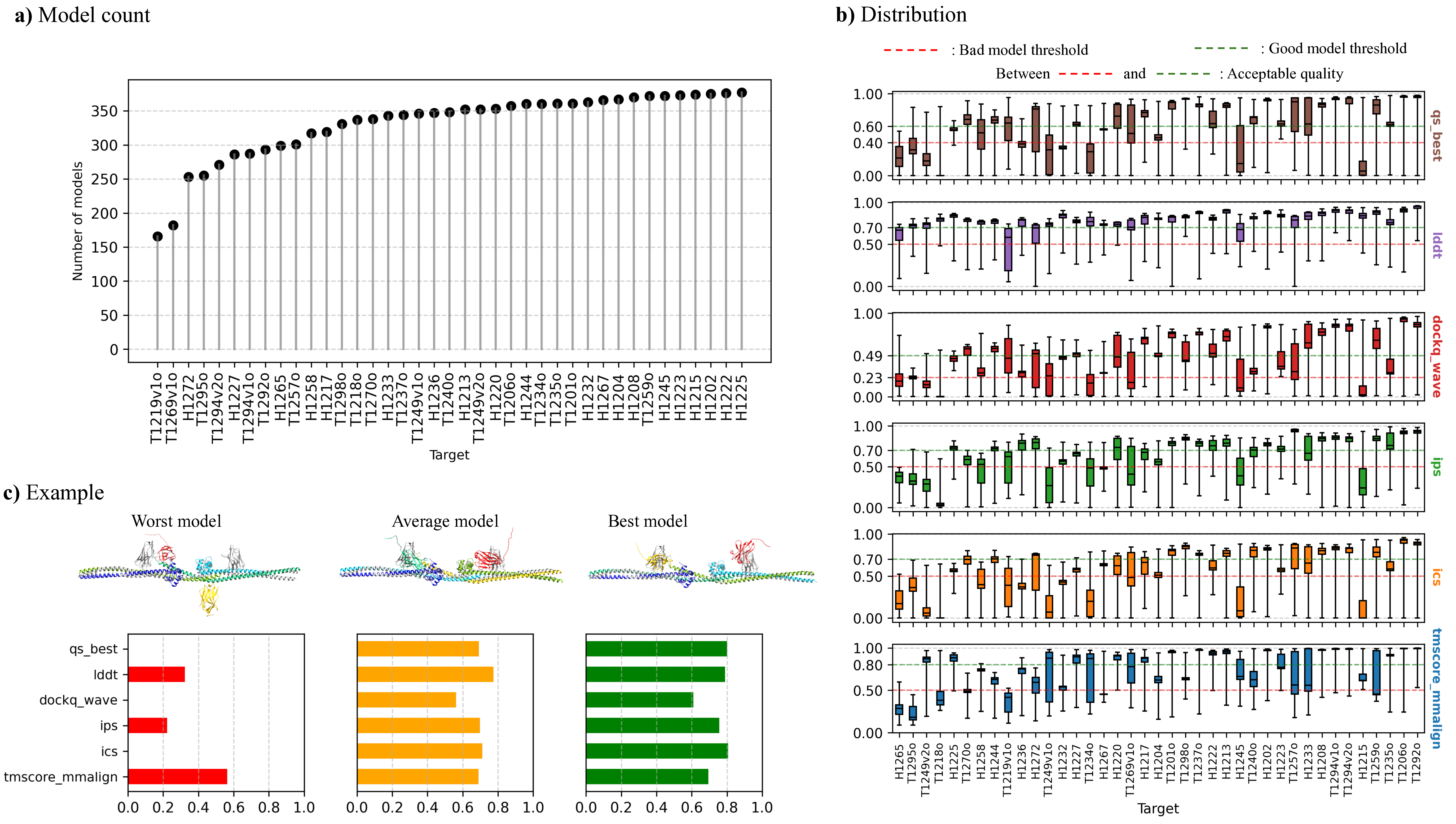}

 \caption{\textbf{CASP16\_community\_dataset.} \textbf{(a) Model count.} Number of models per target in the dataset. \textbf{(b) Score Distribution.} Box plots of each of six representative quality scores of the models for each target. \textbf{(c) Example.} Three representative models (worst, average, best) in terms of sum of the six representative quality scores for a target H1244. Each model with individual chains colored is superimposed with the true structure in gray.}
    
    \label{fig:CASP16_community_dataset}
\end{figure}
\clearpage

\begin{table}[H]
\centering
\caption{Summary of target information and number of models per target for CASP16\_community\_dataset.}
\label{tab:total_models_CASP16_community_dataset}
\small
\begin{tabular}{lllll}
\toprule
  Target &      Stoichiometry &      Protein Classification &  Seq. Length &  Total models \\
\midrule
   H1202 &               A2B2 &           SIGNALING PROTEIN &              380 &               375 \\
   H1204 &             A2B2C2 &            OXYGEN TRANSPORT &              858 &               367 \\
   H1208 &               A1B1 &                     Unknown &              646 &               370 \\
   H1213 &         A1B1C1D1E1 &                     Unknown &             1373 &               352 \\
   H1215 &               A1B1 &                     Unknown &              369 &               374 \\
   H1217 &       A2B2C2D2E2F2 &                     Unknown &             5878 &               319 \\
   H1220 &               A1B4 &                     Unknown &             2515 &               353 \\
   H1222 &             A1B1C1 &                     Unknown &              485 &               376 \\
   H1223 &             A1B1C1 &                     Unknown &              486 &               373 \\
   H1225 &             A1B1C1 &                     Unknown &              483 &               377 \\
   H1227 &               A1B6 &                     Unknown &             5689 &               286 \\
   H1232 &               A2B2 &               VIRAL PROTEIN &              924 &               363 \\
   H1233 &             A2B2C2 & VIRAL PROTEIN/IMMUNE SYSTEM &             1316 &               343 \\
   H1236 &               A3B6 &                       VIRUS &             1929 &               347 \\
   H1244 &             A2B2C2 &                     Unknown &              850 &               360 \\
   H1245 &               A1B1 &                     Unknown &              317 &               372 \\
   H1258 &               A1B2 &                     Unknown &             3092 &               317 \\
   H1265 &              A9B18 &                     Unknown &             3924 &               299 \\
   H1267 &               A2B2 &                     Unknown &             1852 &               366 \\
   H1272 & A1B1C1D1E1F1G1H1I1 &            MEMBRANE PROTEIN &             6879 &               253 \\
  T1201o &                 A2 &           SIGNALING PROTEIN &              420 &               361 \\
  T1206o &                 A2 &               VIRAL PROTEIN &              474 &               357 \\
  T1218o &                 A2 &                     Unknown &             2328 &               337 \\
T1219v1o &                A10 &                     Unknown &              320 &               166 \\
  T1234o &                 A3 &                       VIRUS &             1239 &               360 \\
  T1235o &                 A6 &                       VIRUS &              690 &               361 \\
  T1237o &                 A4 &                     Unknown &             1952 &               344 \\
  T1240o &                 A3 &                     Unknown &             1959 &               348 \\
T1249v1o &                 A3 &                     Unknown &             1464 &               346 \\
T1249v2o &                 A3 &                     Unknown &             1464 &               352 \\
  T1257o &                 A3 &                     Unknown &             3789 &               301 \\
  T1259o &                 A3 &                     Unknown &              729 &               372 \\
T1269v1o &                 A2 &              PROTEIN FIBRIL &             2820 &               182 \\
  T1270o &                 A6 &                     Unknown &             2622 &               338 \\
  T1292o &                 A2 &                     Unknown &              392 &               293 \\
T1294v1o &                 A2 &                     Unknown &              428 &               287 \\
T1294v2o &                 A2 &                     Unknown &              428 &               271 \\
  T1295o &                 A8 &                     Unknown &             3752 &               255 \\
  T1298o &                 A2 &                     Unknown &              684 &               331 \\
\bottomrule
\end{tabular}
\end{table}

\begin{table}[H]
\centering
\caption{Model quality distribution based on dockq\_wave scores (bad: score < 0.23, acceptable: 0.23 <= score < 0.49, good: 0.49 <= score) for CASP16\_community\_dataset.}
\label{tab:dockq_summary_CASP16_community_dataset}
\begin{tabular}{lccc}
\toprule
  Target &  Number of bad models &  Number of acceptable models &  Number of good models \\
\midrule
   H1202 &                     2 &                           37 &                    336 \\
   H1204 &                    21 &                          176 &                    170 \\
   H1208 &                    31 &                           15 &                    324 \\
   H1213 &                    33 &                           16 &                    303 \\
   H1215 &                   333 &                           12 &                     29 \\
   H1217 &                     1 &                           13 &                    305 \\
   H1220 &                    13 &                          174 &                    166 \\
   H1222 &                     2 &                          102 &                    272 \\
   H1223 &                     0 &                          239 &                    134 \\
   H1225 &                     0 &                          289 &                     88 \\
   H1227 &                     5 &                           86 &                    195 \\
   H1232 &                    25 &                          263 &                     75 \\
   H1233 &                    12 &                           35 &                    296 \\
   H1236 &                    86 &                          255 &                      6 \\
   H1244 &                    39 &                           15 &                    306 \\
   H1245 &                   204 &                          104 &                     64 \\
   H1258 &                    53 &                          262 &                      2 \\
   H1265 &                   208 &                           86 &                      5 \\
   H1267 &                    31 &                          331 &                      4 \\
   H1272 &                    69 &                           33 &                    151 \\
  T1201o &                    64 &                            8 &                    289 \\
  T1206o &                    27 &                           30 &                    300 \\
  T1218o &                   286 &                           43 &                      8 \\
T1219v1o &                    32 &                           52 &                     82 \\
  T1234o &                   194 &                          162 &                      4 \\
  T1235o &                    52 &                          238 &                     71 \\
  T1237o &                    16 &                           19 &                    309 \\
  T1240o &                    22 &                          321 &                      5 \\
T1249v1o &                   170 &                          160 &                     16 \\
T1249v2o &                   303 &                           46 &                      3 \\
  T1257o &                    85 &                          112 &                    104 \\
  T1259o &                    11 &                            9 &                    352 \\
T1269v1o &                    94 &                           28 &                     60 \\
  T1270o &                    22 &                           57 &                    259 \\
  T1292o &                     4 &                           11 &                    278 \\
T1294v1o &                     4 &                            3 &                    280 \\
T1294v2o &                    12 &                            3 &                    256 \\
  T1295o &                   117 &                          138 &                      0 \\
  T1298o &                    26 &                          194 &                    111 \\
\bottomrule
\end{tabular}
\end{table}

\subsection{CASP15\_inhouse\_TOP5\_dataset and CASP16\_inhouse\_TOP5\_dataset}
\label{appendix:TOP5_datasets}

CASP15\_inhouse\_TOP5\_dataset and CASP16\_inhouse\_TOP5\_dataset are the subset of CASP15\_inhouse\_dataset and CASP16\_inhouse\_dataset respectively. Each contains only top 5 models for each target predicted by each of dozens of AlphaFold-based predictors in our MULTICOM protein structure prediction system during CASP15 or CASP16, even though each predictor might generate many (e.g., hundreds of) models. These two subsets were used to train and evaluate GATE-AFM. The number of models per target in each of these two subsets is given in the Table \ref{tab:total_models_subsets}.

\begin{table}[h]
\centering
\caption{Summary of the CASP15 and CASP16 in-house TOP5 datasets.}
\label{tab:total_models_subsets}
\begin{tabular}{c|c|c|c}
\toprule
\multicolumn{2}{c|}{\textbf{CASP15\_inhouse\_TOP5\_dataset}} & \multicolumn{2}{c}{\textbf{CASP16\_inhouse\_TOP5\_dataset}} \\

Target & Total Models & Target & Total Models \\
\midrule
H1106 & 290 & H1202 & 390 \\
H1129 & 65 & H1204 & 350 \\
H1134 & 150 & H1208 & 380 \\
H1140 & 130 & H1213 & 300 \\
H1141 & 125 & H1215 & 355 \\
H1142 & 125 & H1220 & 101 \\
H1143 & 115 & H1222 & 345 \\
H1144 & 125 & H1223 & 345 \\
H1151 & 120 & H1225 & 345 \\
H1157 & 115 & H1232 & 250 \\
H1166 & 75 & H1233 & 255 \\
H1167 & 85 & H1236 & 100 \\
H1168 & 75 & H1244 & 330 \\
T1109o & 115 & H1245 & 360 \\
T1110o & 100 & H1258 & 90 \\
T1113o & 210 & H1265 & 77 \\
T1121o & 120 & H1267 & 385 \\
T1123o & 110 & T1201o & 345 \\
T1124o & 100 & T1206o & 325 \\
T1127o & 120 & T1218o & 35 \\
T1132o & 115 & T1219v1o & 305 \\
T1153o & 130 & T1234o & 325 \\
T1160o & 125 & T1235o & 295 \\
T1161o & 125 & T1237o & 275 \\
T1170o & 45 & T1240o & 155 \\
T1173o & 125 & T1257o & 73 \\
T1174o & 95 & T1259o & 305 \\
T1178o & 125 & T1269v1o & 185 \\
T1179o & 125 & T1270o & 290 \\
T1181o & 65 & T1292o & 270 \\
T1187o & 125 & T1295o & 230 \\
 &  & T1298o & 325 \\
\bottomrule
\end{tabular}
\end{table}


\section{Experimental Design}
\label{appendix:experimental_design}

\subsection{Standard EMA Methods in PSBench}
\label{appendix:standard_ema_methods}

PSBench includes six standard EMA methods that are publicly available. They serve as baseline methods for comparison with new EMA methods. Below is a brief overview of each method and its availability:

\begin{itemize}

    \item \textbf{AlphaFold2-Multimer Confidence score (AFM Confidence)}\cite{evans2021protein}: AlphaFold2-Multimer provides self-estimated accuracy estimates for its predicted structures using a confidence score that is computed as a weighted sum of ipTM (interface predicted TM-score) and pTM (predicted TM-score), specifically: 0.8 * ipTM + 0.2 * pTM. This score serves as a single-model EMA method and a strong baseline for datasets generated by AlphaFold2-Multimer or AlphaFold3.
        
    \item \textbf{GATE}\cite{liu2025estimating}: A multi-model EMA approach that leverages graph transformers applied to pairwise similarity graphs derived from input models. GATE combines both single-model and multi-model quality scores from individual models with comparative geometric similarities between models, enabling it to effectively predict the global structural accuracy (e.g., TM-score) of complex structural models. Source code is available at: \url{https://github.com/BioinfoMachineLearning/gate}. \textbf{GATE-AFM}: An enhanced variant of GATE by using AlphaFold2-Multimer features as additional node features. It can be used if such features are available. 
    
    \item \textbf{DProQA}\cite{chen2023gated}: A single-model EMA method based on a Gated Graph Transformer architecture that modulates local neighborhood interactions. It is specifically designed to estimate the interface quality of protein complex models (e.g., DockQ scores) by leveraging a K-nearest neighbor (K-NN) graph representation of the complex structure. Source code is available at: \url{https://github.com/jianlin-cheng/DProQA}.

    \item \textbf{VoroMQA-dark, VoroIF-GNN-score, VoroIF-GNN-pCAD-score}\cite{olechnovivc2023voroif}: A set of single-model EMA methods that utilize the VoroIF-GNN framework (Voronoi Interface Graph Neural Network) to assess protein complex interface quality. These methods operate on Voronoi tessellation-based atomic contact areas, capturing geometric and topological features of the interface. Source code is available at: \url{https://github.com/kliment-olechnovic/ftdmp}.

    \item \textbf{GCPNet-EMA}\cite{morehead2024protein}: An EMA extension of GCPNet (Geometry-Complete Perceptron Network), a deep graph neural network that constructs a 3D graph representation from the atomic point cloud of a protein structure. It predicts both per-residue and per-model structural accuracy estimates, such as local and global lDDT. Although GCPNet-EMA is originally trained on tertiary protein structures (e.g., single-chain models), it can be directly applied to evaluate the accuracy of protein complex structures. Source code is available at: \url{https://github.com/BioinfoMachineLearning/GCPNet-EMA}.
    
    \item \textbf{Average Pairwise Similarity Score (PSS)}\cite{roy2023combining}: A multi-model EMA method that evaluates each predicted complex by computing the average pairwise TM-score between it and all other models in the structural pool using MMalign. This simple yet effective consensus-based approach serves as a strong baseline for estimating the quality of the protein complex model. Source code is available at: \url{https://github.com/BioinfoMachineLearning/MULTICOM_qa}, with a simplified implementation available at: \url{https://github.com/BioinfoMachineLearning/gate/blob/main/gate/feature/mmalign_pairwise.py}.

    
\end{itemize}

It is worth noting that during benchmarking, the quality scores predicted by DProQA, VoroIF-GNN scores, and GCPNet-EMA were normalized by multiplying the raw score by the ratio of the model length to the native structure length. This normalization penalizes shorter decoys, ensuring that the scores more accurately reflect both the completeness and the accuracy of the predicted models relative to the native structures.

\subsection{The Details of Training and Validating GATE}
\label{appendix:detail_training_gate}

GATE and GATE-AFM was first trained, validated, and tested on the CASP15 datasets (i.e., CASP15\_community\_dataset or CASP15\_inhouse\_dataset) respectively and then were blindly evaluated on unseen targets in the CASP16 datasets (i.e., CASP16\_community\_dataset or CASP16\_inhouse\_dataset) during the CASP16 competition from May to August, 2024.

GATE was trained and validated on the CASP15\_community\_dataset, which comprises 10,935 models of 40 protein targets, plus 187 models from another target (e.g., T1115o) whose native structure is not available. For the 10,935 models from 40 protein targets, we used the usalign\_tmscore as labels. For the 187 models without native structure, we obtained the TM-score from \url{https://predictioncenter.org/casp15/multimer_results.cgi?target=T1115o}.
CASP15\_community\_dataset was partitioned into training, validation, and test sets using a 10-fold cross-validation strategy split by protein targets. This dataset was divided into 10 folds, each containing 4–5 targets. For each target, 2,000 subgraphs were sampled from its full model similarity graph, where nodes represent predicted structural models and edges encode pairwise structural similarity. Each subgraph contained up to 50 nodes, resulting in 8,000–10,000 subgraphs per subset. During training, 8 folds were used for parameter optimization, 1 fold for validation (hyperparameter tuning), and 1 fold for testing. This process was repeated 10 times to iterate across all folds. The test targets in each fold are listed in Table~\ref{tab:casp15_official_fold_targets}. The hyperparameter search space is provided in Table~\ref{tab:hyperparameter_search_space}. GATE was trained using a composite loss function combining a pointwise mean squared error (MSE) loss and a pairwise MSE loss, computed over the nodes of each subgraph. The pointwise loss weight was fixed at 1, while the pairwise loss weight was treated as a tunable hyperparameter. To ensure fair model selection, we evaluated trained models on the validation set using the unweighted sum of the pointwise and pairwise losses (i.e., both terms equally weighted) and selected the model with the lowest total loss for testing.

\begin{table}[H]
\caption{Targets assigned to each fold in the CASP15\_community\_dataset for training GATE.}
\centering
\begin{tabular}{lc}
\toprule
  \textbf{Fold} & \textbf{Targets} \\
\midrule
Fold0 & H1135, T1127o, T1161o, T1132o, H1144 \\
Fold1 & H1151, T1153o, H1171, H1114 \\
Fold2 & T1170o, H1166, T1176o, H1134 \\
Fold3 & H1111, H1106, T1109o, T1121o \\
Fold4 & T1174o, T1115o, H1172, H1143 \\
Fold5 & H1137, H1142, T1192o, H1140 \\
Fold6 & T1187o, T1181o, T1179o, T1178o \\
Fold7 & H1168, T1173o, T1160o, H1167 \\
Fold8 & T1113o, H1185, T1123o, H1157 \\
Fold9 & T1110o, H1141, T1124o, H1129 \\
\bottomrule
\end{tabular}
\label{tab:casp15_official_fold_targets}
\end{table}

\begin{table}[H]
\centering
\caption{Hyperparameter search space explored during model fine-tuning.}
\begin{tabular}{ll}
\toprule
\textbf{Hyperparameter} & \textbf{Candidate Values} \\
\midrule
Number of attention heads & 4, 8 \\
Number of graph transformer layers & 2, 3, 4, 5 \\
Dropout rate & 0.1, 0.2, 0.3, 0.4, 0.5 \\
MLP dropout rate & 0.1, 0.2, 0.3, 0.4, 0.5 \\
Hidden dimension & 16, 32, 64 \\
Weight of the pairwise MSE loss & \texttt{auto}, 0.01, 0.05, 0.1, 0.2, 0.3, 0.4, 0.5, 0.6, 0.7, 0.8, 0.9, 1 \\
Optimizer & \texttt{AdamW}, \texttt{SGD} \\
Learning rate & 1e-5, 5e-5, 1e-4, 5e-4, 1e-3 \\
Weight decay & 0.01, 0.05 \\
Layer normalization & False, True \\
Batch size & 256, 400, 512 \\
\bottomrule
\end{tabular}
\label{tab:hyperparameter_search_space}
\end{table}

GATE-AFM was trained and validated by the similar 10-fold cross-validation protocol on CASP15\_inhouse\_TOP5\_dataset. CASP15\_inhouse\_TOP5\_dataset is a subset of CASP15\_inhouse\_dataset, which includes the top 5 ranked models generated by each AlphaFold2-Multimer-based predictor in our in-house MULTICOM3 system. It has 31 protein complex targets. To augment the data, 3,000 subgraphs per target were sampled for each target. In addition to using the node features used in GATE, GATE-AFM incorporates the AlphaFold2-Multimer features as additional node features to enhance prediction accuracy. The test targets in each fold are listed in Table~\ref{tab:casp15_inhouse_fold_targets}. The hyperparameter search space is the same as that used for training GATE and is provided in Table~\ref{tab:hyperparameter_search_space}.
TM-score labels for training GATE\_AFM were generated using an older version of the in-house TM-score calculation script. We provided the TM-scores (tmscore\_usalign\_aligned\_v0) in the CASP15\_inhouse\_dataset for users to reproduce the results if needed. In CASP16, this script was updated to fix minor bugs in residue-residue alignments, with negligible impact on overall score distributions. The updated script is the currently version included in PSBench. 

\begin{table}[H]
\caption{Targets assigned to each fold in the CASP15\_inhouse\_dataset for training GATE-AFM.}
\centering
\begin{tabular}{ll}
\toprule
\textbf{Fold} & \textbf{Targets} \\
\midrule
Fold0 & T1127o, T1161o, T1132o, T1174o \\
Fold1 & T1160o, H1106, H1134 \\
Fold2 & T1173o, T1178o, T1110o \\
Fold3 & T1170o, H1142, H1140 \\
Fold4 & T1179o, T1187o, T1153o \\
Fold5 & T1123o, H1151, T1181o \\
Fold6 & T1121o, T1113o, H1144 \\
Fold7 & H1167, T1124o, H1157 \\
Fold8 & H1168, H1143, H1166 \\
Fold9 & H1141, T1109o, H1129 \\
\bottomrule
\end{tabular}
\label{tab:casp15_inhouse_fold_targets}
\end{table}


\subsection{Ranking of EMA predictors on the CASP16\_community\_dataset based on Z-scores}
\label{appendix:casp16_zscore_ranking}

To assess the overall performance of EMA predictors on the CASP16\_community\_dataset, we ranked the 38 EMA predictors based on cumulative positive Z-scores. For each target, the Z-scores for the predictors were computed separately for the four evaluation metrics (Pearson’s correlation, Spearman’s correlation, ranking loss, and AUROC) based on TM-score. The z-score for a predictor for a target is equal to the original score minus the average score of all the predictors divided by the standard deviation. When calculating total Z-scores for each predictor, only positive Z-scores for each target were accumulated to emphasize strong performances.

In this ranking, MULTICOM\_GATE achieved third place with a total Z-score of 95.7, closely following ModFOLDdock2 (96.9) and MULTICOM\_LLM (105.4).

\begin{figure}[H]
    \centering
    \includegraphics[width=1\linewidth]{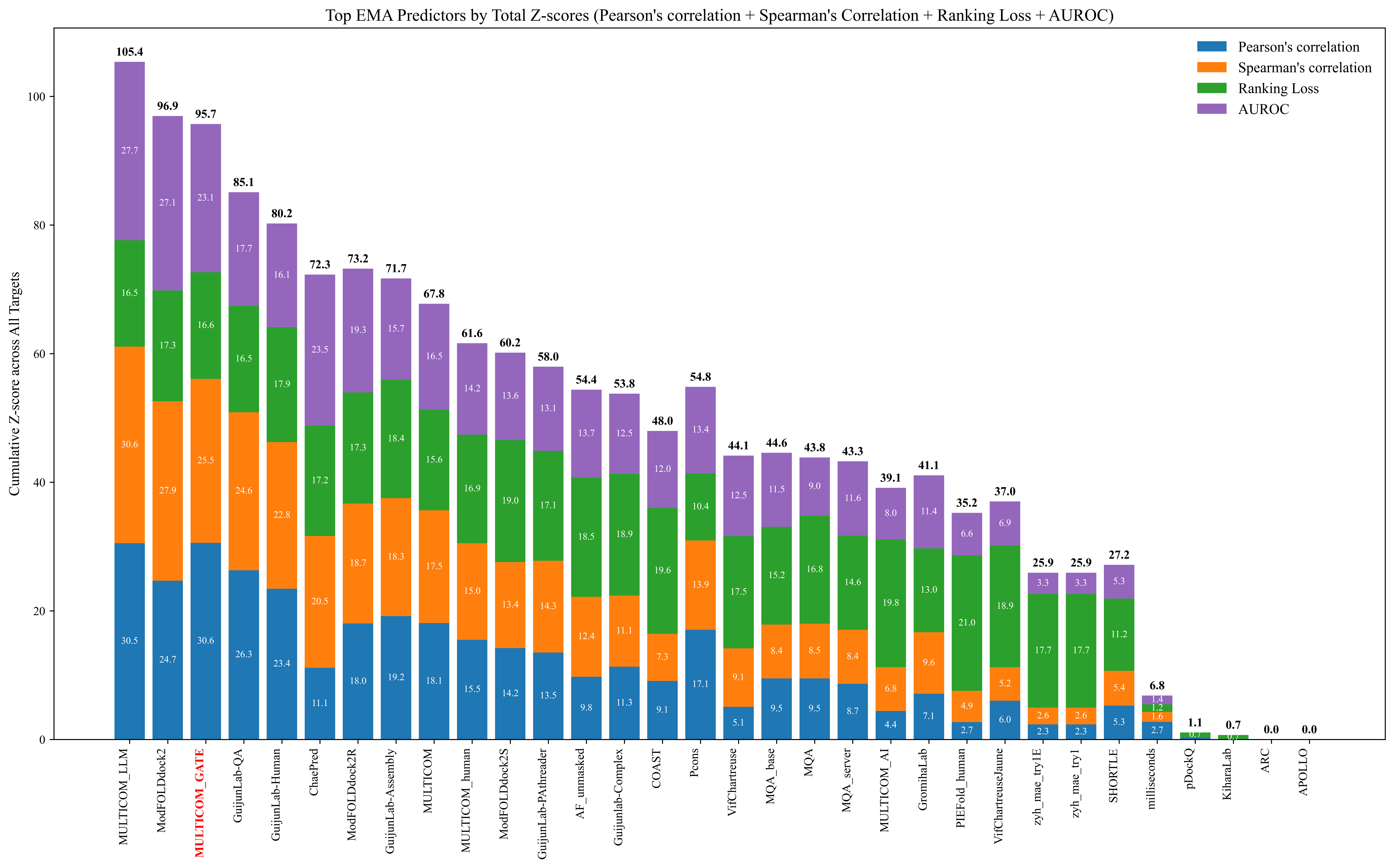}
    \caption{\textbf{CASP16\_community\_dataset results.} Stacked bar plot showing the cumulative positive Z-scores across all CASP16 community targets for the participated EMA predictors. Contributions from four performance metrics (Pearson’s correlation, Spearman’s correlation, ranking loss, and AUROC) are stacked to highlight the overall performance. EMA predictors are ranked by their total Z-scores. MULTICOM\_GATE is colored in red.}
    \label{fig:Z-score_CASP16}
\end{figure}

\section{System Requirements}
\label{appendix:system_requirements}

\subsection{Environment for benchmarking EMA methods with PSBench and labeling model datasets}

The generation of model quality scores and the evaluation of baseline EMA methods were performed on a computing server with the following specifications:

\begin{itemize}
    \item \textbf{Operating System:} CentOS Linux
    \item \textbf{CPU:} AMD EPYC 7552, 3.2 GHz, 48 cores
    \item \textbf{RAM:} 50 GB
\end{itemize}

\subsection{Environment for training and validating GATE}

The training and validation of the GATE and GATE-AFM were conducted on a high-performance computing system with the following configuration:

\begin{itemize}
    \item \textbf{Operating System:} CentOS Linux
    \item \textbf{CPU:} AMD EPYC 7552, 3.2 GHz, 48 cores
    \item \textbf{RAM:} 500 GB
    \item \textbf{GPU:} NVIDIA A100, 80 GB
\end{itemize}

\end{document}